\documentclass[prx,onecolumn,superscriptaddress,longbibliography]{revtex4-2}

\usepackage[]{fontenc}

\usepackage{graphicx}
\usepackage{bm}
\usepackage{mathtools} 
\usepackage{subcaption}
\usepackage{hyperref}
\usepackage{chngcntr} 
\usepackage{multirow}
\usepackage{subeqnarray}
\usepackage{wasysym}

\usepackage[british]{babel}
\usepackage{amsmath}
\usepackage{amssymb}
\usepackage{framed}
\usepackage{soul}
\usepackage{color}
\usepackage{alltt}
\usepackage{pstricks,pst-node,pst-tree,pst-grad,pst-text,graphics}
\usepackage{dcolumn}
\usepackage{url}
\usepackage{cases}
\usepackage{natbib}

\usepackage{ifthen}
\usepackage{afterpage}
\usepackage{comment}
\usepackage{supertabular}
\usepackage{xspace}
\usepackage{setspace}
\usepackage{textcomp}
\usepackage{tikz}
\usepackage{tikz-feynman}
\usetikzlibrary{decorations.pathmorphing}
\usetikzlibrary{decorations.markings}
\usetikzlibrary{arrows,shapes,decorations,automata,backgrounds,petri}

\makeatletter

\newcommand{\creat}[3][]{\@ifempty{#1}{#2^{\dagger}}{\left(#2^{\dagger}\right)^{#1}}\@ifempty{#3}{}{\!(#3)}}
\newcommand{\creatDoi}[3][]{\@ifempty{#1}{\tilde{#2}}{\tilde{#2}^{#1}}\@ifempty{#3}{}{(#3)}}
\newcommand{\annih}[3][]{#2\@ifempty{#1}{}{^{#1}}\@ifempty{#3}{}{(#3)}}

\makeatother

\newcommand{\plaind}{\mathrm{d}}
\newcommand{\dint}[1]{\mathchoice{\!\plaind#1\,}{\!\plaind#1\,}{\!\plaind#1\,}{\!\plaind#1\,}}

\newcommand{\dbar}{\text{\dj}}
\newcommand{\deltabar}{\delta\mkern-6mu\mathchar'26}
\newcommand{\dintbar}[1]{\mathchoice{\!\dbar#1\,}{\!\dbar#1\,}{\!\dbar#1\,}{\!\dbar#1\,}}

\newcommand{\Dint}[1]{\mathcal{D}\!#1\,}
\DeclareFontFamily{U}{wncy}{}
\DeclareFontShape{U}{wncy}{m}{n}{<->wncyr10}{}
\DeclareSymbolFont{mcy}{U}{wncy}{m}{n}
\DeclareMathSymbol{\sha}{\mathord}{mcy}{"58}

\usepackage{dsfont}

\newcommand{\gpset}[1]{\mathds{#1}}

\newcommand{\canetset}[1]{{\mathchoice {\hbox{$\sf\textstyle #1\kern-0.4em #1$}}
{\hbox{$\sf\textstyle #1\kern-0.4em #1$}}
{\hbox{$\sf\scriptstyle #1\kern-0.3em #1$}}
{\hbox{$\sf\scriptscriptstyle #1\kern-0.2em #1$}}}}

\newcommand{\Rset}{\gpset{R}}

\def\nbZ{{\mathchoice {\hbox{$\sf\textstyle Z\kern-0.4em Z$}}
{\hbox{$\sf\textstyle Z\kern-0.4em Z$}}
{\hbox{$\sf\scriptstyle Z\kern-0.3em Z$}}
{\hbox{$\sf\scriptscriptstyle Z\kern-0.2em Z$}}}}

\newcommand{\ident}{\mathbf{1}}

\renewcommand{\AC}{\mathcal{A}}

\newcommand{\phitilde}{\tilde{\phi}}

\newcommand{\psitilde}{\tilde{\psi}}

\renewcommand{\exp}[1]{\mathchoice%
{\mathrm{e}^{#1}}%
{\operatorname{exp}(#1)}
{\operatorname{exp}\left(#1\right)}%
{\operatorname{exp}\left(#1\right)}}

\newcommand{\Eref}[1]{\mbox{Eq.~(\ref{eq:#1})}}
\newcommand{\Erefs}[1]{\mbox{Eqs.~(\ref{eq:#1})}}

\newcommand{\Sref}[1]{Section~\ref{sec:#1}}

\newcommand{\fref}[1]{Fig.~\ref{fig:#1}}

\newlength \standardfigwidth
\DeclareMathAlphabet{\matheub}{U}{eur}{m}{n}

\usepackage{tikz}
\usetikzlibrary{decorations.pathmorphing}
\usetikzlibrary{decorations.markings}
\usetikzlibrary{arrows,shapes,decorations,automata,backgrounds,petri}
\usetikzlibrary{calc}
\usetikzlibrary{patterns}
\usetikzlibrary{decorations.text}

\makeatletter
\newcommand{\ave}[2][]{\@ifempty{#1}{\mathchoice%
{\left\langle #2 \right\rangle}%
{\langle #2\rangle}
{\langle #2\rangle}%
{\langle #2\rangle}}{\mathchoice%
{\left\langle #2 \right\rangle_{#1}}%
{\langle #2\rangle_{#1}}
{\langle #2\rangle_{#1}}%
{\langle #2\rangle_{#1}}}}
\makeatother

\makeatletter
\newcommand{\creatX}[2][]{a^{\@ifempty{#1}{\dagger}{\dagger\,#1}}\@ifempty{#2}{}{(#2)}}
\makeatother
\newcommand{\creatDS}{\tilde{a}}
\makeatletter
\newcommand{\creatDSX}[2][]{\@ifempty{#1}{\creatDS}{\creatDS^{#1}}\@ifempty{#2}{}{(#2)}}
\makeatother
\makeatletter
\newcommand{\annihX}[2][]{a\@ifempty{#1}{}{^{#1}}\@ifempty{#2}{}{(#2)}}
\newcommand{\kernel}[1]{K\@ifempty{#1}{}{\,\!\!\left(#1\right)}}
\newcommand{\kernelDeri}[1]{K'\@ifempty{#1}{}{\left(#1\right)}}

\newcommand{\tableofappendixcontents}{\@starttoc{toa}}
\newcommand{\l@toact}[2]{\@dottedtocline{1}{1.5em}{5.5em}{#1}{#2}}
\newcommand{\l@toactspace}[2]{\ \hfil}
\makeatother

\newcommand{\imag}{\mathring{\imath}}

\renewcommand{\exp}[1]{\mathchoice%
{e^{#1}}%
{\operatorname{exp}(#1)}%
{\operatorname{exp}(#1)}%
{\operatorname{exp}(#1)}}
\newcommand{\corresponding}{\hat{=}}

\usepackage{xspace}

\renewcommand{\st}[1]{}

\newcommand{\av}[1]{\left\langle #1\right\rangle}

\makeatletter
\newcommand{\HMdensity}[2]{\rho_{#2}^{\@ifempty{#1}{}{(#1)}}}
\newcommand{\HMdensityBar}[2]{\overline{\rho}_{#2}^{\@ifempty{#1}{}{(#1)}}}
\newcommand{\HMdensityDot}[2]{\dot{\rho}_{#2}^{\@ifempty{#1}{}{(#1)}}}
\makeatother

\usepackage{textcomp}

\newcommand{\action}{\AC}

\usepackage{color}
\definecolor{darkgreen}{rgb}{0,0.6,0}
\definecolor{darkblue}{rgb}{0,0,0.6}
\definecolor{darkred}{rgb}{0.6,0,0}
\definecolor{darkpurple}{rgb}{0.5,0,0.5}
\hypersetup{
bookmarksopen=true,
pdftitle="",
pdfauthor="", 
pdftoolbar=false, 
pdfstartview={FitH},		
pdfmenubar=true,			
pdfhighlight=/O,			
colorlinks=true,			
urlcolor=darkblue,
citecolor=darkblue,		
linkcolor=darkblue}

\makeatletter
\newcommand{\customlabel}[2]{%
   \protected@write \@auxout {}{\string \newlabel {#1}{{#2}{\thepage}{#2}{#1}{}} }%
   \hypertarget{#1}{\hspace{0pt}}
}
\makeatother

\makeatletter
\newcommand{\product}[3]{\prod_{\@ifempty{#1}{i}{#1} = \@ifempty{#2}{1}{#2}}^{\@ifempty{#3}{N}{#3}}}
\makeatother

\newcommand{\diracdelta}[1]{\delta\left( #1 \right)}



\newcommand{\sdots}{\ifmmode\mathinner{\ldotp\kern-0.2em\ldotp\kern-0.2em\ldotp}\else.\kern-0.13em.\kern-0.13em.\fi}

\newcommand{\selfpropulsion}{\nu_0}
\newcommand{\diffusion}{D}
\newcommand{\transDiffusion}{\diffusion_{x}}
\newcommand{\swaprate}{\alpha}
\newcommand{\telegraphnoise}{w}

\newcommand{\deprate}{\gamma}
\newcommand{\capacity}{n_0}

\newcommand{\masterprob}[1]{\mathcal{P}(#1)}

\makeatletter
\newcommand{\sumoversp}[1]{\sum_{\@ifempty{#1}{s}{#1} \in \{-,+\}}}

\newcommand{\probright}[1]{P_{+}\@ifempty{#1}{}{\left( #1 \right)}}
\newcommand{\probleft}[1]{P_{-}\@ifempty{#1}{}{\left( #1 \right)}}
\newcommand{\probgen}[2]{P_{\@ifempty{#1}{s}{#1}}\@ifempty{#2}{}{\left( #2 \right)}}

\newcommand{\numparticle}[1]{M_{\@ifempty{#1}{s}{#1}}}

\newcommand{\numtracer}[1]{T_{\@ifempty{#1}{s}{#1}}}

\newcommand{\rightparticlefield}[1]{\phi_{+}\@ifempty{#1}{}{\left( #1 \right)}}
\newcommand{\leftparticlefield}[1]{\phi_{-}\@ifempty{#1}{}{\left( #1 \right)}}
\newcommand{\righttracerfield}[1]{\psi_{+}\@ifempty{#1}{}{\left( #1 \right)}}
\newcommand{\lefttracerfield}[1]{\psi_{-}\@ifempty{#1}{}{\left( #1 \right)}}

\newcommand{\particlefield}[2]{\phi_{\@ifempty{#1}{s}{#1}}\@ifempty{#2}{}{\left( #2 \right)}}
\newcommand{\tracerfield}[2]{\psi_{\@ifempty{#1}{s}{#1}}\@ifempty{#2}{}{\left( #2 \right)}}

\newcommand{\rightparticletildefield}[1]{\phitilde_{+}\@ifempty{#1}{}{\left( #1 \right)}}
\newcommand{\leftparticletildefield}[1]{\phitilde_{-}\@ifempty{#1}{}{\left( #1 \right)}}
\newcommand{\righttracertildefield}[1]{\psitilde_{+}\@ifempty{#1}{}{\left( #1 \right)}}
\newcommand{\lefttracertildefield}[1]{\psitilde_{-}\@ifempty{#1}{}{\left( #1 \right)}}

\newcommand{\particlefieldtilde}[2]{\phitilde_{\@ifempty{#1}{s}{#1}}\@ifempty{#2}{}{\left( #2 \right)}}
\newcommand{\tracerfieldtilde}[2]{\psitilde_{\@ifempty{#1}{s}{#1}}\@ifempty{#2}{}{\left( #2 \right)}}

\newcommand{\annihilparticles}[1]{a_{\@ifempty{#1}{s}{#1}}}
\newcommand{\creatparticles}[1]{a_{\@ifempty{#1}{s}{#1}}^{\dagger}}
\newcommand{\annihiltracers}[1]{b_{\@ifempty{#1}{s}{#1}}}
\newcommand{\createtracers}[1]{b_{\@ifempty{#1}{s}{#1}}^{\dagger}}

\makeatother

\newcommand{\actionrnt}{\action_{\rm RnT}}
\newcommand{\actiontracer}{\action_{\rm Tracer}}
\newcommand{\actiontracerint}{\action_{\rm Tracer - int}}
\newcommand{\actiontracerlin}{\action_{\rm Tracer - lin}}

\newcommand{\visitprob}{\mathcal{Q}}
\newcommand{\transfer}{\boldsymbol{T}}
\newcommand{\Pe}{\text{Pe}}

\tikzfeynmanset{
  right particle/.style={
    /tikz/draw=red
  },
  left particle/.style={
    /tikz/draw=blue
  },
  red tracer/.style={
    
    /tikz/draw=red
  },
  left tracer/.style={
    
    /tikz/draw=blue
  },
  right left/.style={
    /tikzfeynman/every edge/.style={/tikzfeynman/.cd, inline={
      \vertex[empty dot] (mid) at ($(i)!0.5!(f)$) {};
      \draw[red] (i) -- (mid);
      \draw[blue] (mid) -- (f);
    }}
  },
  left right/.style=/.style={
    /tikzfeynman/every edge/.style={/tikzfeynman/.cd, inline={
      \vertex[empty dot] (mid) at ($(i)!0.5!(f)$) {};
      \draw[blue] (i) -- (mid);
      \draw[red] (mid) -- (f);
    }}
  }
}

\usetikzlibrary { decorations.pathmorphing, decorations.pathreplacing, decorations.shapes,positioning}

\tikzset{
my dash/.style={thick,dash pattern=on 10pt off 10pt
                }
         }

\tikzset{snake it/.style={decorate, decoration={snake, segment length=1.5mm, amplitude=0.25mm}}}

\def\barerr{\tikz[baseline= - 1.5 ex,  line width=0.6pt]{
  \coordinate (w)  at (0,0);
  \coordinate (mid) at (1,0);
  \coordinate (w0) at (2,0);
  \node[above=0.1em of mid]  {$k,\omega$};
  \draw[red, line width=0.25mm] (w) -- (w0);
}}
\def\barerb{\tikz[baseline= -1.5 ex,  line width=0.6pt]{
  \coordinate (w)    at (0,0);
  \coordinate (wmid) at (1,0);
  \coordinate (w0)   at (2,0);
  \node[above=0.10em of wmid]  {$k,\omega$};
  \draw[red,  line width=0.25mm] (w) -- (wmid);
  \draw[blue, line width=0.25mm] (wmid) -- (w0);
}}
\def\barebb{\tikz[baseline= - 1.5 ex,  line width=0.6pt]{
  \coordinate (w)  at (0,0);
  \coordinate (mid) at (1,0);
  \coordinate (w0) at (2,0);
  \node[above=0.10em of mid]  {$k,\omega$};
  \draw[blue, line width=0.25mm] (w) -- (w0);
}}
\def\barebr{\tikz[baseline= - 1.5 ex,  line width=0.6pt]{
  \coordinate (w)    at (0,0);
  \coordinate (wmid) at (1,0);
  \coordinate (w0)   at (2,0);
  \node[above=0.10em of wmid]  {$k,\omega$};
  \draw[blue,  line width=0.25mm] (w) -- (wmid);
  \draw[red, line width=0.25mm] (wmid) -- (w0);
}}

\def\baretracerr{\tikz[baseline= - 1.5 ex,  line width=0.6pt]{
  \coordinate (w)  at (0,0);
  \coordinate (mid) at (1,0);
  \coordinate (w0) at (2,0);
  \node[above=0.15em of mid] {$k,\omega$};
  \draw[red,line width=0.25mm,snake it] (w) -- (w0);
}}

\def\baretracebb{\tikz[baseline= - 1.5 ex,  line width=0.6pt]{
  \coordinate (w)  at (0,0);
  \coordinate (mid) at (1,0);
  \coordinate (w0) at (2,0);
  \node[above=0.15em of mid] {$k,\omega$};
  \draw[blue,line width=0.25mm,snake it] (w) -- (w0);
}}

\def\transboth{\tikz[baseline= - 0.5 ex,  line width=0.6pt]{
	\fill (0,0) circle (0pt) coordinate (A1);
	\fill (4ex, 0) circle (0pt) coordinate (B1);
    \fill (8ex, 0) circle (0pt) coordinate (C1);
	\draw [color=blue, snake it] (A1)--(B1);
    \draw [color=blue] (B1)--(C1);
    \fill (0,2ex) circle (0pt) coordinate (A2);
	\fill (4ex, 2ex) circle (0pt) coordinate (B2);
    \fill (8ex, 2ex) circle (0pt) coordinate (C2);
	\draw [color=red, snake it] (A2)--(B2);
    \draw [color=red] (B2)--(C2);
}}

\def\barerrnolabel{\tikz[baseline= - 0.5 ex,  line width=0.6pt]{
  \coordinate (w)  at (0,0);
  \coordinate (w0) at (1.5,0);
  \draw[red, line width=0.25mm] (w) -- (w0);
}}
\def\barerbnolabel{\tikz[baseline= - 0.5 ex,  line width=0.6pt]{
  \coordinate (w)    at (0,0);
  \coordinate (wmid) at (0.75,0);
  \coordinate (w0)   at (1.5,0);
  \draw[red,  line width=0.25mm] (w) -- (wmid);
  \draw[blue, line width=0.25mm] (wmid) -- (w0);
}}
\def\barebbnolabel{\tikz[baseline= - 0.5 ex,  line width=0.6pt]{
  \coordinate (w)  at (0,0);
  \coordinate (w0) at (1.5,0);
  \draw[blue, line width=0.25mm] (w) -- (w0);
}}
\def\barebrnolabel{\tikz[baseline= - 0.5 ex,  line width=0.6pt]{
  \coordinate (w)    at (0,0);
  \coordinate (wmid) at (0.75,0);
  \coordinate (w0)   at (1.5,0);
  \draw[blue,  line width=0.25mm] (w) -- (wmid);
  \draw[red, line width=0.25mm] (wmid) -- (w0);
}}

\def\baretracerrnolabel{\tikz[baseline= - 0.5 ex,  line width=0.6pt]{
  \coordinate (w)  at (0,0);
  \coordinate (w0) at (1.5,0);
  \draw[red,line width=0.25mm,snake it] (w) -- (w0);
}}

\def\baretracebbnolabel{\tikz[baseline= - 0.5 ex,  line width=0.6pt]{
  \coordinate (w)  at (0,0);
  \coordinate (w0) at (1.5,0);
  \draw[blue,line width=0.25mm,snake it] (w) -- (w0);
}}

\def\transboth{\tikz[baseline= - 0.5 ex,  line width=0.6pt]{
	\fill (0,0) circle (0pt) coordinate (A1);
	\fill (4ex, 0) circle (0pt) coordinate (B1);
    \fill (8ex, 0) circle (0pt) coordinate (C1);
	\draw [color=blue, snake it] (A1)--(B1);
    \draw [color=blue] (B1)--(C1);
    \fill (0,2ex) circle (0pt) coordinate (A2);
	\fill (4ex, 2ex) circle (0pt) coordinate (B2);
    \fill (8ex, 2ex) circle (0pt) coordinate (C2);
	\draw [color=red, snake it] (A2)--(B2);
    \draw [color=red] (B2)--(C2);
}}

\def\branchboth{\tikz[baseline= - 0.5 ex,  line width=0.6pt]{
	\fill (0,0) circle (0pt) coordinate (A1);
	\fill (4ex, 0) circle (0pt) coordinate (B1);
    \fill (8ex, 0) circle (0pt) coordinate (C1);
    \fill (0,2ex) circle (0pt) coordinate (D1);
	\draw [color=blue] (A1)--(B1);
    \draw [color=blue] (B1)--(C1);
    \draw[color=blue,snake it] (D1)--(B1);
    \fill (0,4ex) circle (0pt) coordinate (A2);
	\fill (4ex, 4ex) circle (0pt) coordinate (B2);
    \fill (8ex, 4ex) circle (0pt) coordinate (C2);
    \fill (0,6ex) circle (0pt) coordinate (D2);
	\draw [color=red] (A2)--(B2);
    \draw [color=red] (B2)--(C2);
    \draw[color=red,snake it] (D2)--(B2);
}}

\def\degradeall{\tikz[baseline= - 0.5 ex,  line width=0.6pt]{
	\fill (0,0) circle (0pt) coordinate (A1);
	\fill (4ex, 0) circle (0pt) coordinate (B1);
    \fill (8ex, 0) circle (0pt) coordinate (C1);
    \fill (8ex,2ex) circle (0pt) coordinate (D1);
	\draw [color=blue] (A1)--(B1);
    \draw [color=blue] (B1)--(C1);
    \draw[color=blue,snake it] (D1)--(B1);
    \fill (0,4ex) circle (0pt) coordinate (A2);
	\fill (4ex, 4ex) circle (0pt) coordinate (B2);
    \fill (8ex, 4ex) circle (0pt) coordinate (C2);
    \fill (8ex,6ex) circle (0pt) coordinate (D2);
	\draw [color=red] (A2)--(B2);
    \draw [color=red] (B2)--(C2);
    \draw[color=red,snake it] (D2)--(B2);
    \fill (10ex,0) circle (0pt) coordinate (A1);
	\fill (14ex, 0) circle (0pt) coordinate (B1);
    \fill (18ex, 0) circle (0pt) coordinate (C1);
    \fill (18ex,2ex) circle (0pt) coordinate (D1);
	\draw [color=blue] (A1)--(B1);
    \draw [color=blue] (B1)--(C1);
    \draw[color=red,snake it] (D1)--(B1);
    \fill (10ex,4ex) circle (0pt) coordinate (A2);
	\fill (14ex, 4ex) circle (0pt) coordinate (B2);
    \fill (18ex, 4ex) circle (0pt) coordinate (C2);
    \fill (18ex,6ex) circle (0pt) coordinate (D2);
	\draw [color=red] (A2)--(B2);
    \draw [color=red] (B2)--(C2);
    \draw[color=blue,snake it] (D2)--(B2);
}}

\def\noiseall{\tikz[baseline= - 0.5 ex,  line width=0.6pt]{
	\fill (0,0) circle (0pt) coordinate (A1);
	\fill (4ex, 0) circle (0pt) coordinate (B1);
    \fill (8ex, 0) circle (0pt) coordinate (C1);
    \fill (8ex,2ex) circle (0pt) coordinate (D1);
    \fill (0ex,2ex) circle (0pt) coordinate (E1);
	\draw [color=blue] (A1)--(B1);
    \draw [color=blue] (B1)--(C1);
    \draw[color=blue,snake it] (D1)--(B1);
    \draw[color=blue,snake it] (E1)--(B1);
    \fill (0,4ex) circle (0pt) coordinate (A2);
	\fill (4ex, 4ex) circle (0pt) coordinate (B2);
    \fill (8ex, 4ex) circle (0pt) coordinate (C2);
    \fill (8ex,6ex) circle (0pt) coordinate (D2);
    \fill (0ex,6ex) circle (0pt) coordinate (E2);
	\draw [color=red] (A2)--(B2);
    \draw [color=red] (B2)--(C2);
    \draw[color=red,snake it] (D2)--(B2);
    \draw[color=red,snake it] (E2)--(B2);
    \fill (10ex,0) circle (0pt) coordinate (A1);
	\fill (14ex, 0) circle (0pt) coordinate (B1);
    \fill (18ex, 0) circle (0pt) coordinate (C1);
    \fill (18ex,2ex) circle (0pt) coordinate (D1);
    \fill (10ex,2ex) circle (0pt) coordinate (E1);
	\draw [color=blue] (A1)--(B1);
    \draw [color=blue] (B1)--(C1);
    \draw[color=red,snake it] (D1)--(B1);
    \draw[color=blue,snake it] (E1)--(B1);
    \fill (10ex,4ex) circle (0pt) coordinate (A2);
	\fill (14ex, 4ex) circle (0pt) coordinate (B2);
    \fill (18ex, 4ex) circle (0pt) coordinate (C2);
    \fill (18ex,6ex) circle (0pt) coordinate (D2);
    \fill (10ex,6ex) circle (0pt) coordinate (E2);
	\draw [color=red] (A2)--(B2);
    \draw [color=red] (B2)--(C2);
    \draw[color=blue,snake it] (D2)--(B2);
    \draw[color=red,snake it] (E2)--(B2);
}}

\def\crazyallone{\tikz[baseline= - 0.5 ex,  line width=0.6pt]{
	\fill (0,0) circle (0pt) coordinate (A1);
    \fill (4ex,2ex) circle (0pt) coordinate (B1);
    \fill (0,4ex) circle (0pt) coordinate (C1);
    \fill (8ex,2ex) circle (0pt) coordinate (D1);
    \fill (8ex,4ex) circle (0pt) coordinate (E1);
    \draw[color=blue,snake it](A1)--(B1);
    \draw[color=blue,snake it](C1)--(B1);
    \draw[color=blue](D1)--(B1);
    \draw[color=blue,snake it](E1)--(B1);
    \fill (0,6ex) circle (0pt) coordinate (A2);
    \fill (4ex,8ex) circle (0pt) coordinate (B2);
    \fill (0,10ex) circle (0pt) coordinate (C2);
    \fill (8ex,8ex) circle (0pt) coordinate (D2);
    \fill (8ex,10ex) circle (0pt) coordinate (E2);
    \draw[color=red,snake it](A2)--(B2);
    \draw[color=red,snake it](C2)--(B2);
    \draw[color=red](D2)--(B2);
    \draw[color=red,snake it](E2)--(B2);

    \fill (10ex,0) circle (0pt) coordinate (A1);
    \fill (14ex,2ex) circle (0pt) coordinate (B1);
    \fill (10ex,4ex) circle (0pt) coordinate (C1);
    \fill (18ex,2ex) circle (0pt) coordinate (D1);
    \fill (18ex,4ex) circle (0pt) coordinate (E1);
    \draw[color=blue,snake it](A1)--(B1);
    \draw[color=red,snake it](C1)--(B1);
    \draw[color=blue](D1)--(B1);
    \draw[color=red,snake it](E1)--(B1);
    \fill (10ex,6ex) circle (0pt) coordinate (A2);
    \fill (14ex,8ex) circle (0pt) coordinate (B2);
    \fill (10ex,10ex) circle (0pt) coordinate (C2);
    \fill (18ex,8ex) circle (0pt) coordinate (D2);
    \fill (18ex,10ex) circle (0pt) coordinate (E2);
    \draw[color=red,snake it](A2)--(B2);
    \draw[color=blue,snake it](C2)--(B2);
    \draw[color=red](D2)--(B2);
    \draw[color=blue,snake it](E2)--(B2);
    }}

\def\crazyalltwo{\tikz[baseline= - 0.5 ex,  line width=0.6pt]{
	\fill (0,0) circle (0pt) coordinate (A1);
    \fill (4ex,2ex) circle (0pt) coordinate (B1);
    \fill (0,4ex) circle (0pt) coordinate (C1);
    \fill (8ex,2ex) circle (0pt) coordinate (D1);
    \fill (8ex,4ex) circle (0pt) coordinate (E1);
    \fill (0,2ex) circle (0pt) coordinate (F1);
    \draw[color=blue](F1)--(B1);
    \draw[color=blue,snake it](A1)--(B1);
    \draw[color=blue,snake it](C1)--(B1);
    \draw[color=blue](D1)--(B1);
    \draw[color=blue,snake it](E1)--(B1);
    \fill (0,6ex) circle (0pt) coordinate (A2);
    \fill (4ex,8ex) circle (0pt) coordinate (B2);
    \fill (0,10ex) circle (0pt) coordinate (C2);
    \fill (8ex,8ex) circle (0pt) coordinate (D2);
    \fill (8ex,10ex) circle (0pt) coordinate (E2);
    \fill (0,8ex) circle (0pt) coordinate (F2);
    \draw[color=red](F2)--(B2);
    \draw[color=red,snake it](A2)--(B2);
    \draw[color=red,snake it](C2)--(B2);
    \draw[color=red](D2)--(B2);
    \draw[color=red,snake it](E2)--(B2);

    \fill (10ex,0) circle (0pt) coordinate (A1);
    \fill (14ex,2ex) circle (0pt) coordinate (B1);
    \fill (10ex,4ex) circle (0pt) coordinate (C1);
    \fill (18ex,2ex) circle (0pt) coordinate (D1);
    \fill (18ex,4ex) circle (0pt) coordinate (E1);
    \fill (10ex,2ex) circle (0pt) coordinate (F1);
    \draw[color=blue](F1)--(B1);
    \draw[color=blue,snake it](A1)--(B1);
    \draw[color=red,snake it](C1)--(B1);
    \draw[color=blue](D1)--(B1);
    \draw[color=red,snake it](E1)--(B1);
    \fill (10ex,6ex) circle (0pt) coordinate (A2);
    \fill (14ex,8ex) circle (0pt) coordinate (B2);
    \fill (10ex,10ex) circle (0pt) coordinate (C2);
    \fill (18ex,8ex) circle (0pt) coordinate (D2);
    \fill (18ex,10ex) circle (0pt) coordinate (E2);
    \fill (10ex,8ex) circle (0pt) coordinate (F2);
    \draw[color=red](F2)--(B2);
    \draw[color=red,snake it](A2)--(B2);
    \draw[color=blue,snake it](C2)--(B2);
    \draw[color=red](D2)--(B2);
    \draw[color=blue,snake it](E2)--(B2);
    }}

\def\obsrr{\tikz[baseline= - 0.5 ex,  line width=0.6pt]{
  \coordinate (w)  at (0,0);
  \coordinate (wmid) at (1,0);
  \coordinate (w0) at (2,0);
  \draw[red, line width=0.25mm, snake it] (w) -- (wmid);
  \draw[red, line width=0.25mm] (wmid) -- (w0);
  \fill[red] (wmid) circle (2.4pt);
}}
\def\obsrb{\tikz[baseline= - 0.5 ex,  line width=0.6pt]{
  \coordinate (w)  at (0,0);
  \coordinate (wmid) at (1,0);
  \coordinate (w0) at (2,0);
  \draw[red, line width=0.25mm, snake it] (w) -- (wmid);
  \draw[blue, line width=0.25mm] (wmid) -- (w0);
  \fill[red] (wmid) circle (2.4pt);
}}
\def\obsbb{\tikz[baseline= - 0.5 ex,  line width=0.6pt]{
  \coordinate (w)  at (0,0);
  \coordinate (wmid) at (1,0);
  \coordinate (w0) at (2,0);
  \draw[blue, line width=0.25mm, snake it] (w) -- (wmid);
  \draw[blue, line width=0.25mm] (wmid) -- (w0);
  \fill[blue] (wmid) circle (2.4pt);
}}
\def\obsbr{\tikz[baseline= - 0.5 ex,  line width=0.6pt]{
  \coordinate (w)  at (0,0);
  \coordinate (wmid) at (1,0);
  \coordinate (w0) at (2,0);
  \draw[blue, line width=0.25mm, snake it] (w) -- (wmid);
  \draw[red, line width=0.25mm] (wmid) -- (w0);
  \fill[blue] (wmid) circle (2.4pt);
}}

\def\obstransrr{\tikz[baseline= - 0.5 ex,  line width=0.6pt]{
  \coordinate (w)  at (0,0);
  \coordinate (wmid) at (1,0);
  \coordinate (wquart) at (1.5,0);
  \coordinate (w0) at (2,0);
  \draw[red, line width=0.25mm, snake it] (w) -- (wmid);
  \draw[blue, line width=0.25mm] (wmid) -- (wquart);
  \draw[red, line width=0.25mm] (wquart) -- (w0);
  \fill[red] (wmid) circle (2.4pt);
}}
\def\obstransrb{\tikz[baseline= - 0.5 ex,  line width=0.6pt]{
  \coordinate (w)  at (0,0);
  \coordinate (wmid) at (1,0);
  \coordinate (wquart) at (1.5,0);
  \coordinate (w0) at (2,0);
  \draw[red, line width=0.25mm, snake it] (w) -- (wmid);
  \draw[red, line width=0.25mm] (wmid) -- (wquart);
  \draw[blue, line width=0.25mm] (wquart) -- (w0);
  \fill[red] (wmid) circle (2.4pt);
}}
\def\obstransbb{\tikz[baseline= - 0.5 ex,  line width=0.6pt]{
  \coordinate (w)  at (0,0);
  \coordinate (wmid) at (1,0);
  \coordinate (wquart) at (1.5,0);
  \coordinate (w0) at (2,0);
  \draw[blue, line width=0.25mm, snake it] (w) -- (wmid);
  \draw[red, line width=0.25mm] (wmid) -- (wquart);
  \draw[blue, line width=0.25mm] (wquart) -- (w0);
  \fill[blue] (wmid) circle (2.4pt);
}}
\def\obstransbr{\tikz[baseline= - 0.5 ex,  line width=0.6pt]{
  \coordinate (w)  at (0,0);
  \coordinate (wmid) at (1,0);
  \coordinate (wquart) at (1.5,0);
  \coordinate (w0) at (2,0);
  \draw[blue, line width=0.25mm, snake it] (w) -- (wmid);
  \draw[blue, line width=0.25mm] (wmid) -- (wquart);
  \draw[red, line width=0.25mm] (wquart) -- (w0);
  \fill[blue] (wmid) circle (2.4pt);
}}

\def\vertrr{\tikz[baseline= - 0.5 ex,  line width=0.6pt]{
  \coordinate (w)  at (0,0);
  \coordinate (wmid) at (0.3,0);
  \coordinate (w0) at (0.6,0);
  \draw[red, line width=0.25mm, snake it] (w) -- (wmid);
  \draw[red, line width=0.25mm] (wmid) -- (w0);
  \fill[red] (wmid) circle (2.4pt);
}}

\def\vertrb{\tikz[baseline= - 0.5 ex,  line width=0.6pt]{
  \coordinate (w)  at (0,0);
  \coordinate (wmid) at (0.3,0);
  \coordinate (w0) at (0.6,0);
  \draw[red, line width=0.25mm, snake it] (w) -- (wmid);
  \draw[blue, line width=0.25mm] (wmid) -- (w0);
  \fill[red] (wmid) circle (2.4pt);
}}

\def\vertbr{\tikz[baseline= - 0.5 ex,  line width=0.6pt]{
  \coordinate (w)  at (0,0);
  \coordinate (wmid) at (0.3,0);
  \coordinate (w0) at (0.6,0);
  \draw[blue, line width=0.25mm, snake it] (w) -- (wmid);
  \draw[red, line width=0.25mm] (wmid) -- (w0);
  \fill[blue] (wmid) circle (2.4pt);
}}

\def\vertbb{\tikz[baseline= - 0.5 ex,  line width=0.6pt]{
  \coordinate (w)  at (0,0);
  \coordinate (wmid) at (0.3,0);
  \coordinate (w0) at (0.6,0);
  \draw[blue, line width=0.25mm, snake it] (w) -- (wmid);
  \draw[blue, line width=0.25mm] (wmid) -- (w0);
  \fill[blue] (wmid) circle (2.4pt);
}}

\def\vertrrzeroth{\tikz[baseline= - 0.5 ex,  line width=0.6pt]{
  \coordinate (w)  at (0,0);
  \coordinate (wmid) at (0.3,0);
  \coordinate (w0) at (0.6,0);
  \draw[red, line width=0.25mm, snake it] (w) -- (wmid);
  \draw[red, line width=0.25mm] (wmid) -- (w0);
}}
\def\vertrrfirst{\tikz[baseline= - 0.5 ex,  line width=0.6pt]{
  \coordinate (w)  at (0,0);
  \coordinate (w1) at (0.3,0);
  \coordinate (w2) at (1.3,0);
  \coordinate (w0) at (1.6,0);
  \draw[red, line width=0.25mm, snake it] (w) -- (w1);
  \draw[red, line width=0.25mm] (w1) -- (w2);
  \draw[red, line width=0.25mm] (w2) -- (w0);
  \draw[red, line width=0.25mm,snake it] (w1) to[out=60, in=120] (w2);
}}
\def\vertrrfirstnoleg{\tikz[baseline= - 0.5 ex,  line width=0.6pt]{
  \coordinate (w1) at (0.3,0);
  \coordinate (w2) at (1.3,0);
  \draw[red, line width=0.25mm] (w1) -- (w2);
  \draw[red, line width=0.25mm,snake it] (w1) to[out=60, in=120] (w2);
}}
\def\vertrrsecondone{\tikz[baseline= - 0.5 ex,  line width=0.6pt]{
  \coordinate (w)  at (0,0);
  \coordinate (w1) at (0.3,0);
  \coordinate (w2) at (1.3,0);
  \coordinate (w3) at (2.3,0);
  \coordinate (w0) at (2.6,0);
  \draw[red, line width=0.25mm, snake it] (w) -- (w1);
  \draw[red, line width=0.25mm] (w1) -- (w2);
  \draw[red, line width=0.25mm] (w2) -- (w3);
  \draw[red, line width=0.25mm] (w3) -- (w0);
  \draw[red, line width=0.25mm,snake it] (w1) to[out=60, in=120] (w2);
  \draw[red, line width=0.25mm,snake it] (w2) to[out=60, in=120] (w3);
}}
\def\vertrrsecondtwo{\tikz[baseline= - 0.5 ex,  line width=0.6pt]{
  \coordinate (w)  at (0,0);
  \coordinate (w1) at (0.3,0);
  \coordinate (wmid1) at (0.8,0);
  \coordinate (w2) at (1.3,0);
  \coordinate (wmid2) at (1.8,0);
  \coordinate (w3) at (2.3,0);
  \coordinate (w0) at (2.6,0);
  \draw[red, line width=0.25mm, snake it] (w) -- (w1);
  \draw[red, line width=0.25mm] (w1) -- (wmid1);
  \draw[blue, line width=0.25mm] (wmid1) -- (w2);
  \draw[blue, line width=0.25mm] (w2) -- (wmid2);
  \draw[red, line width=0.25mm] (wmid2) -- (w3);
  \draw[red, line width=0.25mm] (w3) -- (w0);
  \draw[blue, line width=0.25mm,snake it] (w1) to[out=60, in=120] (w2);
  \draw[red, line width=0.25mm,snake it] (w2) to[out=60, in=120] (w3);
}}

\def\vertrbfirst{\tikz[baseline= - 0.5 ex,  line width=0.6pt]{
  \coordinate (w)  at (0,0);
  \coordinate (w1) at (0.3,0);
  \coordinate (wmid1) at (0.8,0);
  \coordinate (w2) at (1.3,0);
  \coordinate (w0) at (1.6,0);
  \draw[red, line width=0.25mm, snake it] (w) -- (w1);
  \draw[red, line width=0.25mm] (w1) -- (wmid1);
  \draw[blue, line width=0.25mm] (wmid1) -- (w2);
  \draw[blue, line width=0.25mm] (w2) -- (w0);
  \draw[blue, line width=0.25mm,snake it] (w1) to[out=60, in=120] (w2);
}}
\def\vertrbfirstnoleg{\tikz[baseline= - 0.5 ex,  line width=0.6pt]{
  \coordinate (w1) at (0.3,0);
  \coordinate (wmid1) at (0.8,0);
  \coordinate (w2) at (1.3,0);
  \draw[red, line width=0.25mm] (w1) -- (wmid1);
  \draw[blue, line width=0.25mm] (wmid1) -- (w2);
  \draw[blue, line width=0.25mm,snake it] (w1) to[out=60, in=120] (w2);
}}
\def\vertrbsecondone{\tikz[baseline= - 0.5 ex,  line width=0.6pt]{
  \coordinate (w)  at (0,0);
  \coordinate (w1) at (0.3,0);
  \coordinate (wmid1) at (0.8,0);
  \coordinate (w2) at (1.3,0);
  \coordinate (wmid2) at (1.8,0);
  \coordinate (w3) at (2.3,0);
  \coordinate (w0) at (2.6,0);
  \draw[red, line width=0.25mm, snake it] (w) -- (w1);
  \draw[red, line width=0.25mm] (w1) -- (wmid1);
  \draw[blue, line width=0.25mm] (wmid1) -- (w2);
  \draw[blue, line width=0.25mm] (w2) -- (wmid2);
  \draw[blue, line width=0.25mm] (wmid2) -- (w3);
  \draw[blue, line width=0.25mm] (w3) -- (w0);
  \draw[blue, line width=0.25mm,snake it] (w1) to[out=60, in=120] (w2);
  \draw[blue, line width=0.25mm,snake it] (w2) to[out=60, in=120] (w3);
}}
\def\vertrbsecondtwo{\tikz[baseline= - 0.5 ex,  line width=0.6pt]{
  \coordinate (w)  at (0,0);
  \coordinate (w1) at (0.3,0);
  \coordinate (wmid1) at (0.8,0);
  \coordinate (w2) at (1.3,0);
  \coordinate (wmid2) at (1.8,0);
  \coordinate (w3) at (2.3,0);
  \coordinate (w0) at (2.6,0);
  \draw[red, line width=0.25mm, snake it] (w) -- (w1);
  \draw[red, line width=0.25mm] (w1) -- (wmid1);
  \draw[red, line width=0.25mm] (wmid1) -- (w2);
  \draw[red, line width=0.25mm] (w2) -- (wmid2);
  \draw[blue, line width=0.25mm] (wmid2) -- (w3);
  \draw[blue, line width=0.25mm] (w3) -- (w0);
  \draw[red, line width=0.25mm,snake it] (w1) to[out=60, in=120] (w2);
  \draw[blue, line width=0.25mm,snake it] (w2) to[out=60, in=120] (w3);
}}

\def\vertbbzeroth{\tikz[baseline= - 0.5 ex,  line width=0.6pt]{
  \coordinate (w)  at (0,0);
  \coordinate (wmid) at (0.3,0);
  \coordinate (w0) at (0.6,0);
  \draw[blue, line width=0.25mm, snake it] (w) -- (wmid);
  \draw[blue, line width=0.25mm] (wmid) -- (w0);
}}
\def\vertbbfirst{\tikz[baseline= - 0.5 ex,  line width=0.6pt]{
  \coordinate (w)  at (0,0);
  \coordinate (w1) at (0.3,0);
  \coordinate (w2) at (1.3,0);
  \coordinate (w0) at (1.6,0);
  \draw[blue, line width=0.25mm, snake it] (w) -- (w1);
  \draw[blue, line width=0.25mm] (w1) -- (w2);
  \draw[blue, line width=0.25mm] (w2) -- (w0);
  \draw[blue, line width=0.25mm,snake it] (w1) to[out=60, in=120] (w2);
}}
\def\vertbbfirstnoleg{\tikz[baseline= - 0.5 ex,  line width=0.6pt]{
  \coordinate (w1) at (0.3,0);
  \coordinate (w2) at (1.3,0);
  \draw[blue, line width=0.25mm] (w1) -- (w2);
  \draw[blue, line width=0.25mm,snake it] (w1) to[out=60, in=120] (w2);
}}
\def\vertbbsecondone{\tikz[baseline= - 0.5 ex,  line width=0.6pt]{
  \coordinate (w)  at (0,0);
  \coordinate (w1) at (0.3,0);
  \coordinate (w2) at (1.3,0);
  \coordinate (w3) at (2.3,0);
  \coordinate (w0) at (2.6,0);
  \draw[blue, line width=0.25mm, snake it] (w) -- (w1);
  \draw[blue, line width=0.25mm] (w1) -- (w2);
  \draw[blue, line width=0.25mm] (w2) -- (w3);
  \draw[blue, line width=0.25mm] (w3) -- (w0);
  \draw[blue, line width=0.25mm,snake it] (w1) to[out=60, in=120] (w2);
  \draw[blue, line width=0.25mm,snake it] (w2) to[out=60, in=120] (w3);
}}
\def\vertbbsecondtwo{\tikz[baseline= - 0.5 ex,  line width=0.6pt]{
  \coordinate (w)  at (0,0);
  \coordinate (w1) at (0.3,0);
  \coordinate (wmid1) at (0.8,0);
  \coordinate (w2) at (1.3,0);
  \coordinate (wmid2) at (1.8,0);
  \coordinate (w3) at (2.3,0);
  \coordinate (w0) at (2.6,0);
  \draw[blue, line width=0.25mm, snake it] (w) -- (w1);
  \draw[blue, line width=0.25mm] (w1) -- (wmid1);
  \draw[red, line width=0.25mm] (wmid1) -- (w2);
  \draw[red, line width=0.25mm] (w2) -- (wmid2);
  \draw[blue, line width=0.25mm] (wmid2) -- (w3);
  \draw[blue, line width=0.25mm] (w3) -- (w0);
  \draw[red, line width=0.25mm,snake it] (w1) to[out=60, in=120] (w2);
  \draw[blue, line width=0.25mm,snake it] (w2) to[out=60, in=120] (w3);
}}

\def\vertbrfirst{\tikz[baseline= - 0.5 ex,  line width=0.6pt]{
  \coordinate (w)  at (0,0);
  \coordinate (w1) at (0.3,0);
  \coordinate (wmid1) at (0.8,0);
  \coordinate (w2) at (1.3,0);
  \coordinate (w0) at (1.6,0);
  \draw[blue, line width=0.25mm, snake it] (w) -- (w1);
  \draw[blue, line width=0.25mm] (w1) -- (wmid1);
  \draw[red, line width=0.25mm] (wmid1) -- (w2);
  \draw[red, line width=0.25mm] (w2) -- (w0);
  \draw[red, line width=0.25mm,snake it] (w1) to[out=60, in=120] (w2);
}}
\def\vertbrfirstnoleg{\tikz[baseline= - 0.5 ex,  line width=0.6pt]{
  \coordinate (w1) at (0.3,0);
  \coordinate (wmid1) at (0.8,0);
  \coordinate (w2) at (1.3,0);
  \draw[blue, line width=0.25mm] (w1) -- (wmid1);
  \draw[red, line width=0.25mm] (wmid1) -- (w2);
  \draw[red, line width=0.25mm,snake it] (w1) to[out=60, in=120] (w2);
}}
\def\vertbrsecondone{\tikz[baseline= - 0.5 ex,  line width=0.6pt]{
  \coordinate (w)  at (0,0);
  \coordinate (w1) at (0.3,0);
  \coordinate (wmid1) at (0.8,0);
  \coordinate (w2) at (1.3,0);
  \coordinate (wmid2) at (1.8,0);
  \coordinate (w3) at (2.3,0);
  \coordinate (w0) at (2.6,0);
  \draw[blue, line width=0.25mm, snake it] (w) -- (w1);
  \draw[blue, line width=0.25mm] (w1) -- (wmid1);
  \draw[red, line width=0.25mm] (wmid1) -- (w2);
  \draw[red, line width=0.25mm] (w2) -- (wmid2);
  \draw[red, line width=0.25mm] (wmid2) -- (w3);
  \draw[red, line width=0.25mm] (w3) -- (w0);
  \draw[red, line width=0.25mm,snake it] (w1) to[out=60, in=120] (w2);
  \draw[red, line width=0.25mm,snake it] (w2) to[out=60, in=120] (w3);
}}

\def\vertbrsecondtwo{\tikz[baseline= - 0.5 ex,  line width=0.6pt]{
  \coordinate (w)  at (0,0);
  \coordinate (w1) at (0.3,0);
  \coordinate (wmid1) at (0.8,0);
  \coordinate (w2) at (1.3,0);
  \coordinate (wmid2) at (1.8,0);
  \coordinate (w3) at (2.3,0);
  \coordinate (w0) at (2.6,0);
  \draw[blue, line width=0.25mm, snake it] (w) -- (w1);
  \draw[blue, line width=0.25mm] (w1) -- (wmid1);
  \draw[blue, line width=0.25mm] (wmid1) -- (w2);
  \draw[blue, line width=0.25mm] (w2) -- (wmid2);
  \draw[red, line width=0.25mm] (wmid2) -- (w3);
  \draw[red, line width=0.25mm] (w3) -- (w0);
  \draw[blue, line width=0.25mm,snake it] (w1) to[out=60, in=120] (w2);
  \draw[red, line width=0.25mm,snake it] (w2) to[out=60, in=120] (w3);
}}

\begin{document}
\newcommand{\titleText}{Run, Tumble and Paint}

\title{\titleText}

\author{Emir Sezik}
\thanks{These authors contributed equally}
\email{emir.sezik19@imperial.ac.uk}

\author{Callum Britton}
\thanks{These authors contributed equally}
\email{callum.britton19@imperial.ac.uk}

\author{Alex Touma}
\email{alex.touma24@imperial.ac.uk}

\author{Gunnar Pruessner}
\email{g.pruessner@imperial.ac.uk}

\affiliation{%
Department of Mathematics
and Centre of Complexity Science, 
Imperial College London, London SW7 2AZ, United Kingdom}%

\date{today}

\begin{abstract}
The visit probability, quantifying whether a particle has reached a given point for the first time by a specified time, provides access to various extreme value statistics and serves as a fundamental tool for characterising active matter models. However, previous studies have largely neglected how the visit probability depends on the internal degree of freedom driving the active particle. To address this, we calculate the ``state-dependent'' visit probability for a Run-and-Tumble particle, that is the probability that the particle first passes through $x$ before time $t$, keeping track of its internal state during first passage. This process may be thought of as the particle ``painting'' the positions it passes through for the time in the colour of its self-propulsion state. We perform this calculation in one dimension using Doi-Peliti field theory, by extending the tracer mechanism from previous works to incorporate such ``polar deposition'' and demonstrate that state-dependent visit probabilities can be elegantly captured within this field-theoretic framework. We further derive the total volume covered by a right- (or left-) moving Run-and-Tumble particle and compare our results with known expressions for Brownian motion.
\end{abstract}

\maketitle

\section{Introduction}
Active matter systems, where particles consume energy to exert local force typically in the form of self-propulsion \cite{vrugt2025exactlyactivematter}, have attracted much interest from the statistical mechanics community for its relevance in the description of living systems \cite{Hydro_soft_Active}. The development of tools and methods to analyse such systems is key to understand emergent complex behaviour ranging from flocking \cite{Novel_pt_vicsek, LRO_toner} to motility induced phase separation \cite{stat_mech_RNT,cates_motility-induced_2015}. Paradigmatic models such as Run-and-Tumble (RnT) \cite{RNT_confining}, where a particle moves in approximately straight runs interrupted by Poissonian tumbles that reorientate the particle, has been shown to capture some of the dynamics of motile bacteria like \textit{Escherichia coli} (E. coli) \cite{deGennes2004, Control_of_RnT}. These models typically consist of endowing a Brownian particle with an additional, typically persistent, degree of freedom that undergoes a stochastic evolution and is coupled uni-directionally to the position \cite{stat_mech_RNT, ABP_golestanian, AOUP}. Such a coupling drives the system out of equilibrium and generally produce non-Gaussian statistics, rendering the analysis of the particle's motion, through the calculation of survival probabilities, first-passage properties and extreme statistics \cite{RNT_confining, malakarSteadyStateRelaxation2018a, current_fluctuations}, non-trivial. Exact results are limited to a few models, e.g. \cite{ABP2dBasu, ABPkumar, RNT_confining, malakarSteadyStateRelaxation2018a} and one typically needs to resort to a perturbative expansion \cite{FPT_perturbative} in order to make meaningful progress. 

Doi-Peliti field theory \cite{peliti_path_1985, doi_second_1976}, a perturbative scheme, has been used with great success to characterise the probability densities of active matter models \cite{zhangFieldTheoryFree2022, Britton_2025, garcia-millan_run-and-tumble_2021, DP_AOUP}. By mapping the model onto a field theory, it allows for the employment of an efficient perturbative expansion organised by diagrams. Though perturbative, the field theory is capable of producing \emph{exact} results, demonstrating the versatility of this method. Doi-Peliti field theory has also been shown to be suitable to characterise the extreme statisics of both Markovian and, more crucially, non-Markovian systems \cite{walterFieldTheorySurvival2022}. In particular, by considering a parallel process where the particles deposit immobile particles at the locations they visit \cite{nekovarFieldtheoreticApproachWiener2016}, the field theory provides access to first passage statistics and survival probabilities. 

While first passage properties, survival probabilities, and extreme value statistics of active matter models has received considerable attention, the dependence of such quantities on the internal degree of freedom has been largely neglected. Keeping track of the internal degree of freedom allows for a more comprehensive understanding of their statistics. Beyond the characterisation of active matter models, such ``state-dependent'' statistics has been proven to be essential for the control of active matter \cite{szilard_chen, szilard_malgaretti, PhysRevLett.131.188301}. Specifically, some of us recently argued Ref.~\cite{sezik2025conditionalsplittingprobabilitieshiddenstate} that keeping track of the internal degree of freedom in such statistics can be used to infer the potentially hidden internal state of an active particle upon exit events, which can be used in the construction of active information engines \cite{neri2025pre, cocconi2025scipost}.

Motivated by the lack of results regarding state-dependent statistics, in this work, we calculate the ``state-dependent'' visit probability of an RnT particle, defined as the probability with which the particle passes through $x$ for the first time as a right, or as a left, mover by time $t$, given it was initialised at position $x_0$ as a right, or as a left, mover at time $t_0$ using Doi-Peliti field theory. This extends the definition of the visit probability to processes with multiple degrees of freedom, allowing for a more comprehensive characterisation of their statistics. To keep track of the internal degree of freedom within a field-theoretic framework for calculating state-dependent visit probabilities, we extend the tracer mechanism proposed in Ref. \cite{nekovarFieldtheoreticApproachWiener2016} to allow for a polar deposition. Now, an RnT particle deposits tracers of a species corresponding to the self-propulsion orientation of the particle at the time when it visited that point in space for the first time. The notion of polar deposition is ubiquitous in biological and synthetic active matter, where agents modify their environments based on their instantaneous polarity. Examples include migrating cells depositing extracellular matrix (ECM) fibres as polarity \cite{Discher2005}
, ant pheromone trails that encode directional bias \cite{Deneubourg1990} 
and chemical self-phoretic colloids that leave behind orientation-dependent chemical-gradients \cite{Anderson1989}.
The implementation of a polar tracer mechanism thus not only furthers our ability to analyse out-of-equilibrium theoretical models, but also provides a stepping stone in understanding complex biological motility.


The paper is structured as follows: In \Sref{model} we introduce polar tracer deposition mechanism. This is followed by the derivation of a Doi-Peliti field theory for such a particle in \Sref{ft}, in particular deriving the Doi-Peliti action, bare propagators and perturbative vertices that enable the calculation of key observables. In \Sref{observables}, we focus on the calculation of the \emph{state-dependent} visit probabilities, that is the probability of the RnT particle first visiting some point in space $x$ with a particular orientation given a particular initialisation. This allows the calculation of the long-time total volume explored in each state, and the total volume explored in each half plane in each respective state. We conclude and discuss in \Sref{conc}.

\section{Model}\label{sec:model}
A one-dimensional RnT particle, with position $x(t)$, obeys the Langevin equation
\begin{equation} \label{eq:RnT_Langevin}
    \dot{x}(t) = \selfpropulsion \telegraphnoise(t) + \sqrt{2 \transDiffusion} \xi(t)
\end{equation}
where $\transDiffusion$ is the diffusion constant and $\xi(t)$ is a unit Gaussian noise satisfying:
\begin{equation}
    \av{ \xi(t)} = 0, ~~ \av{\xi(t) \xi(t')} = \delta(t-t').
\end{equation}
The self-propulsion is captured in \Eref{RnT_Langevin} by the velocity $\selfpropulsion$ and the internal self-propulsion state or orientation $\telegraphnoise(t) = \{-1,1\}$, which is a telegraph noise with switching rate $\swaprate/2$, or tumbling rate $\alpha$. The sign of the telegraph noise $\telegraphnoise(t)$ determines the direction of the drift of the particle. For $\telegraphnoise(t) = 1$, the particle is called a right- mover as it moves, in addition to its diffusion, ballistically to the right, whereas for $\telegraphnoise(t) = -1$, the particle is called a left-mover as it ballistically moves to the left in addition to diffusion. We denote the orientation of the particle by $s\in \{-,+\}$, where $s = -$ corresponds to a left-moving particle and $s = +$ to a right-moving particle. Eq.~\eqref{eq:RnT_Langevin} can be recast into a Fokker-Planck equation for $\probright{x,t}$ and $\probleft{x,t}$, the probability densities of finding a right and a left moving particle at position $x$ at time $t$, respectively:
\begin{subequations} \label{eq:FPE_RnT}
    \begin{align}
        \frac{\partial \probright{x,t}}{\partial t} &= \transDiffusion \frac{\partial^2 \probright{x,t}}{\partial x^2} - \selfpropulsion \frac{\partial \probright{x,t}}{\partial x} + \frac{\swaprate}{2}\left(\probleft{x,t} - \probright{x,t} \right), \label{eq:FPE_RnT_right}\\
        \frac{\partial \probleft{x,t}}{\partial t} &= \transDiffusion \frac{\partial^2 \probleft{x,t}}{\partial x^2} + \selfpropulsion \frac{\partial \probleft{x,t}}{\partial x} + \frac{\swaprate}{2}\left(\probright{x,t} - \probleft{x,t} \right), \label{eq:FPE_RnT_left}
    \end{align}
\end{subequations}
where the tumble rate, $\alpha$, is twice the switching rate of the telegraph process as a particle can tumble onto the direction in which it's moving \cite{zhangFieldTheoryFree2022}.  

We are interested in the \emph{state-dependent} visit probabilities of run-and-tumble particles. 
This visit probability, $\visitprob(x,s, t;x_0,s_0,t_0)$, is defined as the probability that the particle has passed through $x$ before time $t$ and was in state $s$ upon first passage, given it was initialised at position $x_0$ and internal state $s_0$ at time $t_0$.
%
In  Refs.~\cite{walterFieldTheorySurvival2022,nekovarFieldtheoreticApproachWiener2016}, it was shown that visit probabilities can be computed in field theory by defining an additional process 
whereby the particle deposits immobile tracers at every point it visits for the first time. This process,``the tracer mechanism'', can be used to keep track of where the particle has been. To keep track of the state-dependent visit probability, we choose a deposition mechanism such that the RnT particle leaves a tracer of a particular species $s \in \{-,+\}$ that reflects the RnT particle's orientation as it visits a position \emph{for the first time}. As only ever one tracer is placed at every position visited, the pattern of tracers is a record of the RnT particle's orientation during first passages of positions. As illustrated in \fref{1}, this process can be thought of as the RnT particle ``painting" the real line. Once a region is painted, it cannot be further painted over as the paint mark at a given point captures the particle's first passage through that point. Without diffusion, the resulting spatial pattern is deterministic: All positions to the right of the initial condition carry tracers of species $s = +$ and all positions to the left carry $s= - $. Diffusion, however, produces islands of the opposite species of size $\sim \transDiffusion/\selfpropulsion$ within stretches of length $\sim \selfpropulsion/\swaprate$, producing non-trivial statistics for the paint marks. To capture the tracer mechanism within Doi-Peliti field theory, we consider the occupation numbers $M_s$ of RnT particles and $T_s$ of the tracer particles at any point in space. The tracer mechanism is then defined by the following transitions
\begin{subequations} \label{eq:deposition_rates}
    \begin{align}
        \numparticle{}  &\longrightarrow \numparticle{} + \numtracer{}~~ \text{with rate}~~ \deprate\left(1 - \frac{1}{\capacity}\sumoversp{} \numtracer{} \right), \label{eq:deposition}\\
        \numtracer{} &\longrightarrow \emptyset, ~~ \text{with rate} ~~\varepsilon \label{eq:extinction}
    \end{align}
\end{subequations}
where the deposition of the tracers is governed by rate $\deprate$ times an amplitude that vanishes linearly as the total local occupation $T_{+} + T_{-}$ reaches carrying capacity $n_0$. Specifically, choosing $\capacity = 1$ will allow us to have at most one tracer particle per position. That $n_0$ merely features as a scale factor is a result of the limit $\deprate \to \infty$ to be taken to guarantee tracer deposition at every passage \cite{walterFieldTheorySurvival2022, nekovarFieldtheoreticApproachWiener2016}. Alternative schemes, \cite{wijlandFieldTheoryReactiondiffusion2001}, seem to be less suitable for the present perturbative scheme. We have endowed the tracers with an infinitesimal extinction rate, $\varepsilon$, that will be taken to $0$ at the end of any calculation. On the basis of Eq.~\eqref{eq:deposition_rates}, we can write down the temporal evolution of the probability $\masterprob{\numparticle{}, \numtracer{}}$ to find $M_s$ RnT particles and $T_s$ tracers at a given position, for all process governing the tracers
\begin{align}\label{eq:Master_Eq_deposition}
    \frac{d\masterprob{\numparticle{}, \numtracer{};t}}{dt}&=\varepsilon \sumoversp{} \left[ (\numtracer{} + 1)\masterprob{\numparticle{}, \numtracer{} + 1;t} - \numtracer{}  \masterprob{\numparticle{}, \numtracer{};t}\right] \nonumber \\
    & +\deprate \sumoversp{s} \numparticle{} \left[\left(1 - \frac{\sumoversp{s'} \numtracer{s'} -1}{\capacity}\right)\masterprob{\numparticle{}, \numtracer{}-1;t} - \left(1 - \frac{\sumoversp{s'} \numtracer{s'} }{\capacity}\right)\masterprob{\numparticle{}, \numtracer{};t} \right].
\end{align}

\begin{figure}[t!]
\centering
\includegraphics[width=0.75\textwidth]{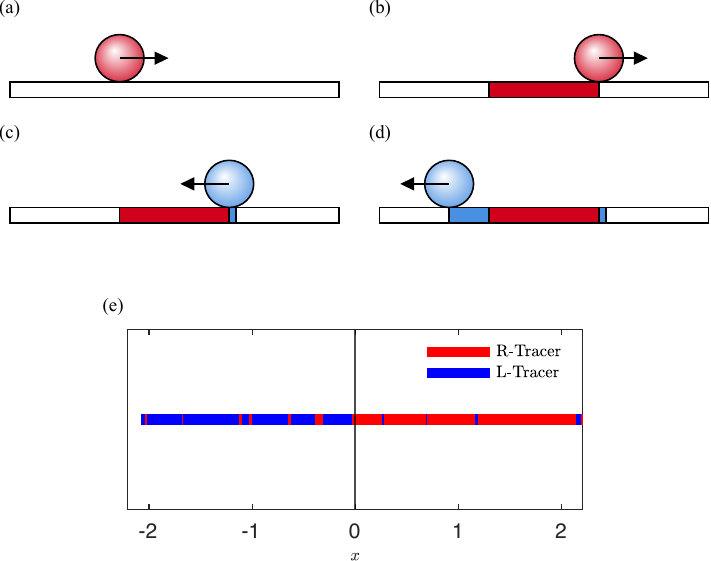}
\caption{The tracer mechanism --- Cartoon: (a) The RnT particle, shown as a ball, is initialised as a right mover (red). (b) Particle evolves in time, travelling on average to the right and leaving a trail of right (red) tracers, or painting the real line with a colour corresponding to its orientation. (c) Particle tumbles to become a left mover (blue) and the jitter of its diffusive motion leaves a ``puddle" of blue to the right of the red stretch. (d) Particle travels further to the left and leaves left (blue) tracers \textit{specifically} where the particle has not already traversed. The particle doesn't paint over the existing red tracers. (e) Numerical realisation of the process, initialised as a right mover with $\swaprate = 10.0;\, \transDiffusion = 1.0;\, \selfpropulsion = 1.0;\, T=100$. }
\label{fig:1}
\end{figure}
\section{Field Theory}\label{sec:ft}
The Master Eq.~\eqref{eq:Master_Eq_deposition} and the Fokker-Planck Eqs. \eqref{eq:FPE_RnT} can be recast into a Doi-Peliti field theory defined by an action $\action = \actionrnt + \actiontracer$ \cite{tauber_critical_2014, zhangFieldTheoryFree2022, walterFieldTheorySurvival2022,pruessnerFieldTheoriesActive2025a}, that can be split into a bilinear part containing the RnT dynamics
\begin{equation}\label{Eq:RnT_action}
\actionrnt = \int \dint{x} \int \dint{t} \left\{\sumoversp{} \particlefieldtilde{}{x,t}\left( \partial_t - \transDiffusion \partial^2_x +s\selfpropulsion\partial_x + \frac{\swaprate}{2} \right)\particlefield{}{x,t} - \frac{\swaprate}{2}\sumoversp{} \particlefieldtilde{}{x,t}\particlefield{-s}{x,t}\right\}
\end{equation}
and a part containing the tracer deposition mechanism, $\actiontracer$, which can further be decomposed into a bilinear and a perturbative part $\actiontracer = \actiontracerlin + \actiontracerint$
\begin{align} \label{Eq:Tracer_actionlin}
    \actiontracerlin &=  \int \dint{x} \int \dint{t} \sumoversp{}\tracerfieldtilde{}{x,t}\left( \partial_t + \varepsilon\right) \tracerfield{}{x,t}\\
    \label{Eq:Tracer_actionint}\actiontracerint &= - \deprate \int \dint{x} \int \dint{t} \sumoversp{s}\left( \particlefieldtilde{}{x,t}+1 \right) \particlefield{}{x,t} \tracerfieldtilde{}{x,t}\left(1 - \frac{1}{\capacity}\sumoversp{s'}\left( \tracerfieldtilde{s'}{x,t} + 1\right) \tracerfield{s'}{x,t} \right).
\end{align}
The perturbative expansion can be carried out by expanding the right most exponential in the path integral for a given observable
\begin{equation}
    \av{\bullet} = \int \Dint{\phi} \Dint{\phitilde} \Dint{\psi} \Dint{\psitilde}\exp{-\action} \bullet= \int \Dint{\phi} \Dint{\phitilde} \Dint{\psi} \Dint{\psitilde}\exp{-\actionrnt - \actiontracerlin} \bullet \exp{-\actiontracerint}.
\end{equation}
From $\actionrnt$ in Eq.~\eqref{Eq:RnT_action}, we identify the bare propagators corresponding to the dynamics of an RnT particle, which read in Fourier-space and -time as 
\begin{multline}\label{eq:particle_bareprop}
    \begin{pmatrix}
    \ave[0]{\particlefield{+}{k,\omega}\particlefieldtilde{+}{k',\omega'}}&&\ave[0]{\particlefield{+}{k,\omega}\particlefieldtilde{-}{k',\omega'}}\\\ \ave[0]{\particlefield{-}{k,\omega}\particlefieldtilde{+}{k',\omega'}}&&\ave[0]{\particlefield{-}{k,\omega}\particlefieldtilde{-}{k',\omega'}}
    \end{pmatrix}
    \\
    =\frac{\deltabar(k+k')\deltabar(\omega+\omega')}{(-\imag\omega+\transDiffusion k^2)(-\imag\omega+\transDiffusion k^2+\swaprate) + \selfpropulsion^2k^2}\begin{pmatrix} -\imag\omega -\imag\selfpropulsion k + \transDiffusion k^2 +\swaprate/2 && \swaprate/2\\ \swaprate/2 && -\imag\omega +\imag\selfpropulsion k + \transDiffusion k^2 +\swaprate/2 \end{pmatrix}\\
    \corresponding \deltabar(k+k')\deltabar(\omega+\omega')
    \begin{pmatrix}
    \barerr && \barerb \\
    \barebr && \barebb
\end{pmatrix}
.
\end{multline}
To maintain causality, we may introduce a positive mass $\epsilon \to 0^{+}$ where necessary, such that the poles of the propagators come to lie in the lower half-plane. We will not always make this technical detail explicit in the following. Once transformed back to real space and time, the propagators quantify the probability density of finding an RnT particle for any time at a particular position and in a particular self-propulsion state given it was initialised at a certain position and with a certain orientation at a previous time. The elements of the matrix in Eq.~\eqref{eq:particle_bareprop} account for all tumbling events. The diagonal elements therefore have contributions only from even numbers of orientation switches and the off-diagonal elements only from odd numbers.

From $\actiontracerlin$, Eq.\eqref{Eq:Tracer_actionlin}, we identify bare propagators of the tracer fields
\begin{equation}\label{eq:tracer_bareprop}
\begin{aligned}
    \begin{pmatrix}
    \ave[0]{\tracerfield{+}{k,\omega}\tracerfieldtilde{+}{k',\omega'}}&&\ave[0]{\tracerfield{+}{k,\omega}\tracerfieldtilde{-}{k',\omega'}}\\\ \ave[0]{\tracerfield{-}{k,\omega}\tracerfieldtilde{+}{k',\omega'}}&&\ave[0]{\tracerfield{-}{k,\omega}\tracerfieldtilde{-}{k',\omega'}}
    \end{pmatrix}
    &= \frac{\deltabar(k+k')\deltabar(\omega+\omega')}{-\imag\omega+\varepsilon}\ident_2 \\
    &\corresponding
    \deltabar(k+k')\deltabar(\omega+\omega')
    \begin{pmatrix}
    \baretracerr && 0\\
    0 && \baretracebb
\end{pmatrix},
\end{aligned}
\end{equation}
where $\ident_2$ is the $2\times2$ identity matrix. The diagonal elements for both species of tracers are identical as the tracer particles are immobile and serve the sole purpose of recording the orientation of the RnT particle as it passes for the first time by a point. The off-diagonal elements are $0$ as there is no bare process in which a right tracer transmutes to a left tracer or vice versa. Additionally, we identify the amputated perturbative vertices from $\actiontracerint$, Eq.\eqref{Eq:Tracer_actionint}, which are all related to the deposition of tracers
\begin{equation}\label{eq:vertices} 
\begin{aligned}
\actiontracerint = \int \dint{x} \int \dint{t} &\bigl[\overbrace{-\deprate\sum_s \tracerfieldtilde{s}{}\particlefield{s}{}}^{\transboth} + \overbrace{\deprate\sum_s \tracerfield{s}{}\particlefieldtilde{s}{}\particlefield{s}{}}^{\branchboth} + \overbrace{\frac{\deprate}{n_0}\sum_{s,s'} \tracerfieldtilde{s}{}\particlefield{s}{}\tracerfield{s'}{}}^{\degradeall} + \overbrace{\frac{\deprate}{n_0}\sum_{s,s'} \tracerfieldtilde{s}{}\particlefieldtilde{s}{}\particlefield{s}{}\tracerfield{s'}{}}^{\noiseall}\nonumber\\&+\underbrace{\frac{\deprate}{n_0}\sum_{s,s'}\tracerfieldtilde{s}{}\tracerfieldtilde{s'}{}\particlefield{s}{}\tracerfield{s'}{}}_{\crazyallone}+\underbrace{\frac{\deprate}{n_0}\sum_{s,s'}\tracerfieldtilde{s}{}\tracerfieldtilde{s'}{}\particlefieldtilde{s}{}\particlefield{s}{}\tracerfield{s'}{}}_{\crazyalltwo}\bigr].
\end{aligned}
\end{equation}
There is a structure in this plethora of vertices. Specifically, the colour of incoming and outgoing straight lines, if both present, must be identical, as the RnT particle do not re-orient during the process of deposition. Furthermore, as far as our calculation is concerned, we will need to use only a select number of perturbative vertices. 

\section{Observables}\label{sec:observables}
In this section, we compute the state-dependent visit probability, $\visitprob(x,s,t; x_0, s_0,t_0)$, namely the probability that an RnT particle, initialised at $x_0$ and with orientation $s_0$ at time $t_0$ has by time $t$ passed a point $x$ for the first time with orientation $s$. 
In other words, we calculate the usual visit probability \cite{Cox1970} of $x$ before time $t$ of a particle and further ask what self-propulsion state the particle was in when it passed through $x$ for the first time. 
%
Within our field-theoretic framework, we probe whether at time $t$ there is a tracer particle of species $s$ present at position $x$
\begin{equation} \label{eq:observable_defn}
    \visitprob(x,s,t; x_0, s_0,t_0) \equiv \lim_{\deprate \to \infty} \frac{1}{n_0} \av{\tracerfield{s}{x,t} \particlefieldtilde{s_0}{x_0,t_0}}
\end{equation}
where $\particlefieldtilde{s_0}{x_0,t_0}$ is the Doi-shifted particle-creator field, initialising an RnT particle of orientation $s_0$ at $x_0$ and time $t_0$ and $\tracerfield{s}{x,t}$ is the tracer-annihilator field, measuring the number density of tracer particles of orientation $s$ at position $x$ and time $t$. The division by the carrying capacity density $n_0$ ensures that the resulting quantity is a probability \cite{walterFieldTheorySurvival2022} and the limit $\deprate \to \infty$ renders the track of tracers continuous, capturing the full path of the RnT particle. The limit $\deprate \to \infty$ needs to be taken after the perturbative expansion has been resummed to avoid any perturbative artefacts. In what follows, we set $x_0 = 0$ and $t_0 = 0$ such that the state-dependent visit probability can be written succinctly as 
\begin{equation} \label{eq:observable_defn_simplified}
    \visitprob_{s,s'}(x,t) \equiv  \visitprob(x,s,t; 0, s',0)
\end{equation}

To simplify the convolutions due to consecutive Poisson processes, we work in the Fourier basis, turning convolutions into products. Therefore, we first aim for the observable $\av{\tracerfield{s}{k,\omega} \particlefieldtilde{s'}{k',\omega'}}$. Diagrammatically, we can write the field-theoretic observable as
\begin{multline}\label{eq:observable_formal}
    \begin{pmatrix}
    \ave{\tracerfield{+}{k,\omega}\particlefieldtilde{+}{k',\omega'}}&&\ave{\tracerfield{+}{k,\omega}\particlefieldtilde{-}{k',\omega'}}\\\ \ave{\tracerfield{-}{k,\omega}\particlefieldtilde{+}{k',\omega'}}&&\ave{\tracerfield{-}{k,\omega}\particlefieldtilde{-}{k',\omega'}}
    \end{pmatrix}  \\
    \corresponding\,\deltabar(k+k') \deltabar(\omega+\omega') \begin{pmatrix}
    \obsrr + \obstransrr && \obsrb + \obstransrb \\
    \obsbr + \obstransbr && \obsbb + \obstransbb
\end{pmatrix}
\end{multline}
The appearance of the Dirac $\delta-$functions, enforcing $k+k'= 0$ and $\omega+\omega'= 0$, implements the spatial and temporal translational invariance of the field theory. It is convenient to factor them out, so that the diagrams depend on a single wavevector $k$ and a single frequency $\omega$. In Eq.~\eqref{eq:observable_formal}, the filled bullets in the diagrams contain all contributions due to the vertices on Eq.~\eqref{eq:vertices} to all orders of the perturbative expansion. While they are connected to the external fields by the bare propagators in Eqs.~\eqref{eq:particle_bareprop} and \eqref{eq:tracer_bareprop}, internally they may contain loops formed by any of the terms in Eq. \eqref{eq:vertices}. The challenge of the present field theory is to account for all such terms systematically. As a first step, we rewrite Eq.~\eqref{eq:observable_formal} as a matrix multiplication
\begin{multline} \label{eq:observable_matmul}
    \begin{pmatrix}
    \ave{\tracerfield{+}{k,\omega}\particlefieldtilde{+}{k',\omega'}}&&\ave{\tracerfield{+}{k,\omega}\particlefieldtilde{-}{k',\omega'}}\\\ \ave{\tracerfield{-}{k,\omega}\particlefieldtilde{+}{k',\omega'}}&&\ave{\tracerfield{-}{k,\omega}\particlefieldtilde{-}{k',\omega'}}
    \end{pmatrix}  \\
    \corresponding\, \deltabar(k+k') \deltabar(\omega+\omega')\begin{pmatrix}
    \baretracerrnolabel && 0 \\
    0 && \baretracebbnolabel
\end{pmatrix}\begin{pmatrix}
    \vertrr && \vertrb \\
    \vertbr && \vertbb
\end{pmatrix} \begin{pmatrix}
    \barerrnolabel && \barerbnolabel \\
    \barebrnolabel && \barebbnolabel
\end{pmatrix}
\end{multline}
where the middle matrix on the right contains all vertices where a straight line of the orientation indicated by the colour enters and a wavy line of the orientation indicated by the colour leaves. Replacing the diagrams by suitable algebraic terms in their Fourier-representation, means we will be able to use standard matrix multiplication to calculate Eq.~\eqref{eq:observable_matmul}. The amputated diagrams contributing to the filled bullets are given by:
\begin{subequations} \label{eq:loops}
\begin{alignat}{3}
    \vertrr &=& \vertrrzeroth &+& \vertrrfirst + \vertrrsecondone + \vertrrsecondtwo  + \dots \label{eq:loops_rr}\\
    \vertrb &=& \makebox[\widthof{$\vertrrzeroth$}][c]{0} &+& \vertrbfirst + \vertrbsecondone + \vertrbsecondtwo + \dots  \label{eq:loops_rb}\\
    \vertbr &=& \makebox[\widthof{$\vertrrzeroth$}][c]{0} &+& \vertbrfirst   + \vertbrsecondone + \vertbrsecondtwo   + \dots \label{eq:loops_br}\\
    \vertbb &=& \vertbbzeroth &+& \vertbbfirst   + \vertbbsecondone   + \vertbbsecondtwo + \dots \label{eq:loops_bb}
\end{alignat}
\end{subequations}
where we have included contributions only up to $2-$loops. There are no tree-level contributions to the \emph{vertex} of an RnT particle of a particular orientation depositing a tracer of the opposite species, producing $0$ in Eqs.~\eqref{eq:loops_rb} and \eqref{eq:loops_br}, even when the correlation function contain such off-diagonal terms, Eq. \eqref{eq:observable_formal}, by the RnT particle undergoing a transmutation first. The loop diagrams, in Eqs.~\eqref{eq:loops}, capture the events where an RnT particle revisits a point. Specifically, the $n-$loop diagrams order correspond to a particle \emph{returning} to the same position for the $n^{\rm th}$ time. Accounting for such returns corrects lower order terms as the higher order diagrams correspond to the reduced (or vanishing) deposition rate when the RnT particle attempts to leave a tracer on a site it has previously visited. Ignoring the species of the stumps of the amputated diagrams on the right of Eqs.~\eqref{eq:loops}, which will be accounted for in Eq.~\eqref{eq:geometric_structure_algebra} by the correct couplings, we may think of the chain of loops as being generated in a matrix multiplication
\begin{equation} \label{eq:geometric_structure}
    \begin{pmatrix}
    \vertrr && \vertrb \\
    \vertbr && \vertbb
\end{pmatrix} = \begin{pmatrix}
    \vertrrzeroth && 0 \\
    0 && \vertbbzeroth
\end{pmatrix} + \begin{pmatrix}
    \vertrrfirst && \vertrbfirst \\
    \vertbrfirst && \vertbbfirst
\end{pmatrix} + \begin{pmatrix}
    \vertrrfirst && \vertrbfirst \\
    \vertbrfirst && \vertbbfirst
\end{pmatrix}^{2} + \dots
\end{equation}
Such a structure will hold to all orders in the perturbative expansion such that the full vertices are given by the geometric series:
\begin{align} \label{eq:geometric_structure_algebra}
    \begin{pmatrix}
    \vertrr && \vertrb \\
    \vertbr && \vertbb
    \end{pmatrix}
    &= \deprate \ident_{2} - \frac{\deprate^2}{\capacity} \transfer(\omega) + \frac{\deprate^3}{\capacity^2}\transfer^2(\omega) + \dots = \gamma  \sum_{n = 0}^ {\infty} \left(-\frac{\transfer(\omega)}{\capacity}\right)^n = \deprate\left(\ident_2+\frac{\deprate}{\capacity} \transfer(\omega)\right)^{-1}
\end{align}
where $\transfer(\omega)$ is the matrix comprised of the loop diagrams in Eq.~\eqref{eq:geometric_structure}
\begin{align} 
    \transfer(\omega) &= \begin{pmatrix}
    \vertrrfirstnoleg && \vertrbfirstnoleg \\
    \vertbrfirstnoleg && \vertbbfirstnoleg
    \end{pmatrix} \nonumber  \\
    &= \int\dintbar{k'} \int \dintbar{\omega'}\frac{1}{\imag(\omega + \omega') + \varepsilon}\begin{pmatrix} \frac{-\imag\omega' -\imag\selfpropulsion k' + \transDiffusion k'^2 +\swaprate/2}{(-\imag\omega'+\transDiffusion k'^2)(-\imag\omega'+\transDiffusion k'^2+\swaprate) + \selfpropulsion^2k'^2} && \frac{\swaprate/2}{(-\imag\omega'+\transDiffusion k'^2)(-\imag\omega'+\transDiffusion k'^2+\swaprate) + \selfpropulsion^2k'^2}\\ \frac{\swaprate/2}{(-\imag\omega'+\transDiffusion k'^2)(-\imag\omega'+\transDiffusion k'^2+\swaprate) + \selfpropulsion^2k'^2} && \frac{-\imag\omega' +\imag\selfpropulsion k' + \transDiffusion k'^2 +\swaprate/2}{(-\imag\omega'+\transDiffusion k'^2)(-\imag\omega'+\transDiffusion k'^2+\swaprate) + \selfpropulsion^2k'^2} \end{pmatrix} \label{eq:transfer_matrix_integral}\\
    &= \frac{1}{2\transDiffusion\sqrt{-\imag \omega (-\imag \omega + \swaprate)}(\sqrt{\beta_{+}} + \sqrt{\beta_{-}})} \begin{pmatrix} -\imag \omega + \swaprate/2 + \sqrt{-\imag \omega(-\imag \omega + \swaprate)} && \swaprate/2\\ \swaprate/2 &&  -\imag \omega + \swaprate/2 + \sqrt{-\imag \omega(-\imag \omega + \swaprate)} \end{pmatrix} \label{eq:transfer_matrix_result}.
\end{align}
Going from Eq.~\eqref{eq:transfer_matrix_integral} to Eq. \eqref{eq:transfer_matrix_result},  we have carried out the integrals over $\omega'$ and $k'$ using the roots
\begin{equation}  \label{eq:betafactordefn}
    \beta_{\pm} = \frac{-\imag \omega + \swaprate/2 + \selfpropulsion^2/(2 \transDiffusion)}{\transDiffusion} \pm \frac{\sqrt{(-\imag \omega + \swaprate/2 + \selfpropulsion^2/(2\transDiffusion))^2 + \imag \omega(-\imag \omega + \swaprate)}}{\transDiffusion}.
\end{equation}
That $\transfer(\omega)$ is symmetric and has the same diagonal elements reflects that only the relative orientation of the RnT particle and the tracer is relevant in quantifying the return processes to a single point. 

\begin{figure}
     \centering
     \begin{subfigure}[b]{0.48\textwidth}
         \centering
         \includegraphics[width=\textwidth]{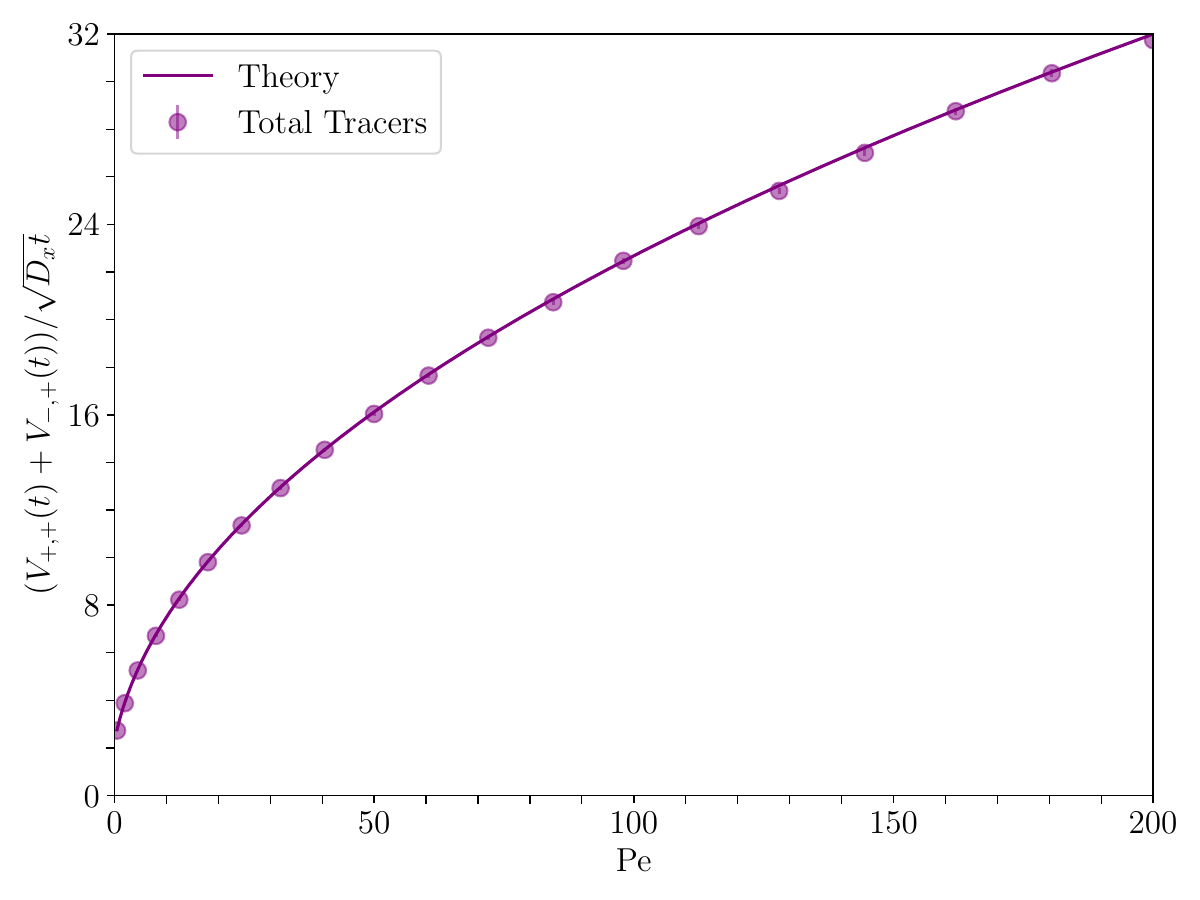}
         \caption{}
         \label{fig:2a}
     \end{subfigure}
     \begin{subfigure}[b]{0.48\textwidth}
         \centering
         \includegraphics[width=\textwidth]{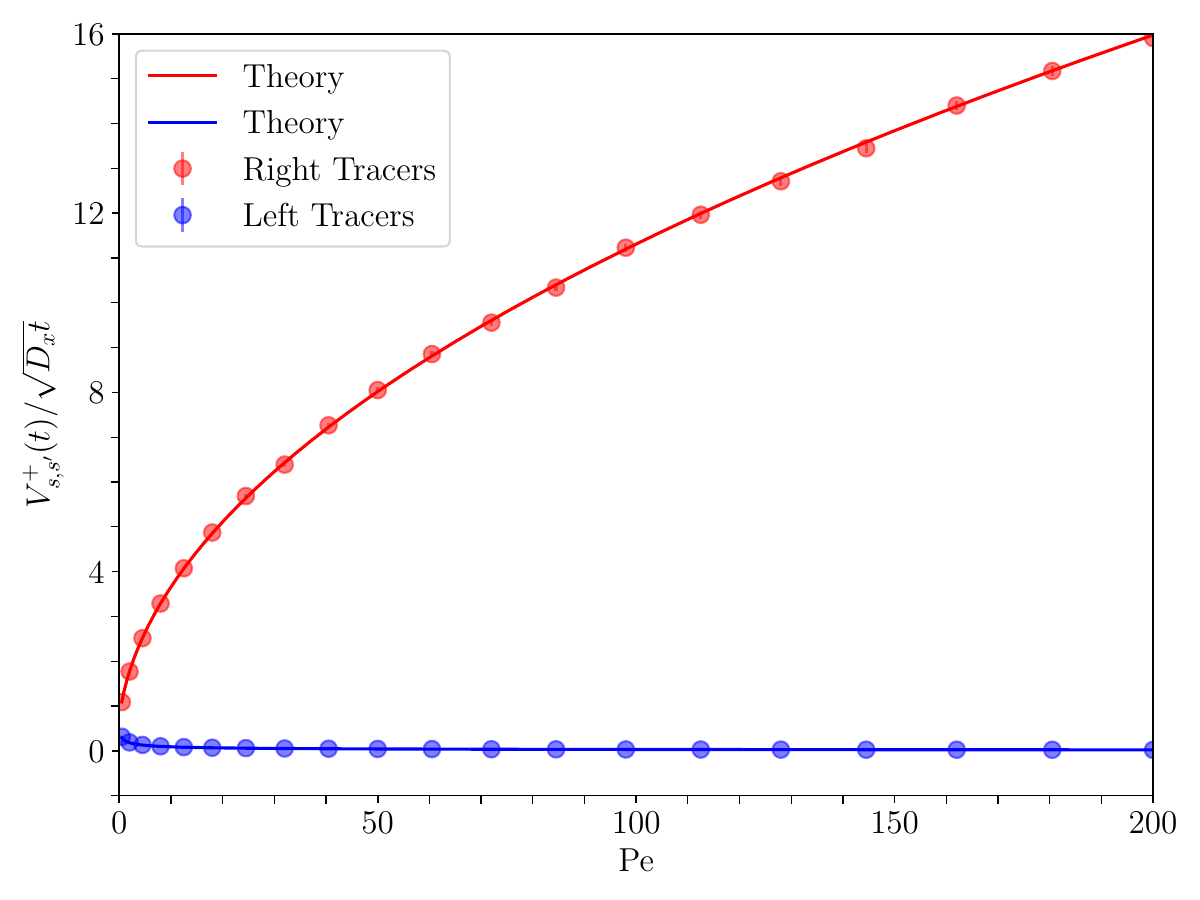}
         \caption{}
          \label{fig:2b}
     \end{subfigure}
        \caption{Comparison with simulations of (a) the asymptotic expression for the total volume covered by an RnT particle, Eq.~\eqref{eq:total_Volume_covered} and (b) the volume covered on the positive half line in each state as a function of Péclet, Eq.~\eqref{eq:asymptotic_volume_explored_half_line} with a fixed large time, $t \gg 1/\swaprate$, with $1/\swaprate$ the tumbling timescale. For convenience, the vertical axis is scaled by $\sqrt{\transDiffusion t}$ to elucidate the dependence on the Péclet number, $\Pe$. The error bars associated with the numerical results are shown, but are smaller than the symbol size. For simplicity, we simulate the system on a lattice in discrete time and count all the distinct lattice sites visited during time $t \gg 1/\swaprate$ to obtain our observables. Simulations were run for $10^{5}$ time steps, with ensemble averaging over $10^{4}$, for $D_x=1.0$; $\alpha=0.5$; $\nu_0\in\{0.5, 1.0,\dots,10.0\}$.}
        \label{fig:simul_comparison}
\end{figure}

Taking the limit $\deprate \to \infty$ in Eq.~\eqref{eq:geometric_structure_algebra} reduces the matrix of dressed vertices in Eq.~\eqref{eq:geometric_structure_algebra} to $\capacity \transfer^{-1}(\omega)$. Using these vertices in Eq.~\eqref{eq:observable_matmul}, together with the propagators in Eqs.~\eqref{eq:tracer_bareprop} and \eqref{eq:particle_bareprop}, we can rewrite the desired observable in Eq.~\eqref{eq:observable_defn} and thus Eq.~\eqref{eq:observable_defn_simplified} as
\begin{multline} \label{eq:visit_prob_fourier}
    \bm{\visitprob}(k,\omega) = \begin{pmatrix}
        \visitprob_{++}(k,\omega) & \visitprob_{+-}(k,\omega)\\
        \visitprob_{-+}(k,\omega) &  \visitprob_{--}(k,\omega)
    \end{pmatrix}
    \\=
    \frac{1}{-\imag \omega + \varepsilon} \transfer^{-1}(\omega) \cdot 
    \begin{pmatrix} \frac{-\imag\omega -\imag\selfpropulsion k + \transDiffusion k^2 +\swaprate/2}{(-\imag\omega'+\transDiffusion k^2)(-\imag\omega'+\transDiffusion k^2+\swaprate) + \selfpropulsion^2k^2} && \frac{\swaprate/2}{(-\imag\omega+\transDiffusion k^2)(-\imag\omega+\transDiffusion k^2+\swaprate) + \selfpropulsion^2k^2}\\ \frac{\swaprate/2}{(-\imag\omega+\transDiffusion k^2)(-\imag\omega+\transDiffusion k^2+\swaprate) + \selfpropulsion^2k^2} && \frac{-\imag\omega +\imag\selfpropulsion k + \transDiffusion k^2 +\swaprate/2}{(-\imag\omega+\transDiffusion k^2)(-\imag\omega+\transDiffusion k^2+\swaprate) + \selfpropulsion^2k^2} \end{pmatrix}
\end{multline}
Eq.~\eqref{eq:visit_prob_fourier} is the key result of the present section. Inverse Fourier transforming it in time and space would yield the full state-dependent visit probability of a particle that was initialised with orientation $s$ at $x_0 = 0$ and $t_0 = 0$. Using Eq.~\eqref{eq:visit_prob_fourier}, we can also extract the expected range or distinct volume $V_{s,s'}(t)$ covered for the first time by time $t$ by a particle with orientation $s$ given it was initialised as a particle with orientation $s'$ at time $t_0 = 0$, which is simply the integral over all space of the state-dependent visit probability $\visitprob_{s,s'}(x,t)$:
\begin{equation}\label{eq:volume_def}
    V_{s,s'}(t) = \int\dint{x} \visitprob_{s,s'}(x,t) = \int \dintbar{\omega} e^{-\imag \omega t} \visitprob_{s,s'}(k = 0, \omega).
\end{equation}
At $k = 0$, Eq.~\eqref{eq:visit_prob_fourier} simplifies to
\begin{equation} \label{eq:volume_explored_fourier}
    \bm{\visitprob}(k = 0, \omega) = \transDiffusion
    \begin{pmatrix}
        \frac{\sqrt{\beta_+} + \sqrt{\beta_{-}} }{-\imag \omega\sqrt{-\imag \omega(-\imag \omega +\swaprate)}} & -\frac{\alpha\left(\sqrt{\beta_+} + \sqrt{\beta_{-}}\right) }{\omega \left(\alpha(2\omega + \imag \sqrt{-\imag \omega(-\imag \omega +\alpha)} ) + 2\omega(-\imag \omega + \sqrt{-\imag \omega(-\imag \omega +\swaprate)}) \right)}  \\
        -\frac{\alpha\left(\sqrt{\beta_+} + \sqrt{\beta_{-}}\right) }{\omega \left(\alpha(2\omega + \imag \sqrt{-\imag \omega(-\imag \omega +\alpha)} ) + 2\omega(-\imag \omega + \sqrt{-\imag \omega(-\imag \omega +\swaprate)}) \right)} & \frac{\sqrt{\beta_+} + \sqrt{\beta_{-}}}{-\imag \omega\sqrt{-\imag \omega(-\imag \omega +\swaprate)}}
    \end{pmatrix}
\end{equation}
It is not feasible to perform the inverse Fourier transform of Eq.~\eqref{eq:volume_explored_fourier} to obtain $V_{s,s'}(t)$ for all times. Some limiting cases are relegated to Appendix \ref{app:limiting_cases}. However, we can find the asymptotic behaviour of $V_{s,s'}(t)$ for large $t$ by focusing on the $\omega-$pole close to the origin. After determining the residue of $\visitprob(k = 0, \omega)$ at $\omega = 0$, the volume $V_{s,s'}(t)$, for any $\swaprate>0$, becomes:
\begin{multline} \label{eq:asymptotic_volume_explored}
    \boldsymbol{V}(t) = \begin{pmatrix}
        V_{++}(t) & V_{+-}(t)\\
        V_{-+}(t) &  V_{--}(t)
    \end{pmatrix} \\= \sqrt{\transDiffusion }\sqrt{1 + \Pe} \lim_{\epsilon \to 0^{+}}\int \dintbar{\omega} \frac{\exp{-\imag \omega t}}{(-\imag \omega + \epsilon)^{3/2}} \begin{pmatrix}
        1 & 1\\
        1 & 1 
    \end{pmatrix}
    + \mathcal{O}(t^0)
    = \frac{\sqrt{\transDiffusion t}}{\Gamma(3/2)} \sqrt{1 + \Pe} \begin{pmatrix}
        1 & 1\\
        1 & 1 
    \end{pmatrix}
    + \mathcal{O}(t^0).
\end{multline}
where we have used the integral 
\begin{equation} \label{eq:integral_identity}
    \lim_{\epsilon \to 0^{+}} \int \dintbar{\omega}\frac{\exp{-\imag \omega t}}{(-\imag \omega + \epsilon)^{a}} = \frac{t^{a-1}}{\Gamma(a)} \theta(t)
\end{equation}
based on the Schwinger trick \cite{schwinger_trick}, $x^{-a} = \int_0^{\infty} \dint{s} s^{a-1} \exp{-s x}/\Gamma(a)$ with $x = -\imag \omega +\epsilon$. In Eq.~\eqref{eq:asymptotic_volume_explored}, we have introduced the Péclet number $\Pe = \selfpropulsion^2/(\alpha \transDiffusion)$, quantifying the relative strength of activity to thermal motion. As expected, the asymptotic form of the volume explored is independent of the initial orientation of the particle and the same for a left and right-moving particle. For long enough times, the transient excursion of the particle due to its initial orientation represents a tiny fraction of the total volume covered. For $\alpha t \gg 1$, the total distinct volume visited is dominated by 
\begin{equation} \label{eq:total_Volume_covered}
    V_{+,s}(t) + V_{-,s}(t) = 4\sqrt{\frac{\transDiffusion (1+\Pe) t}{\pi}} + \mathcal{O}(t^0),
\end{equation}
identical to the leading order Wiener-sausage volume in one dimension for a Brownian particle with diffusivity $\transDiffusion (1+\Pe)$ \cite{nekovarFieldtheoreticApproachWiener2016}. The factor $\transDiffusion (1+\Pe)$ also characterises the mean-squared displacement of an RnT particle \cite{zhangFieldTheoryFree2022}, corresponding to the effective diffusivity of the particle. In Fig.~\ref{fig:2a}, we show simulation results for $V_{+,s}(t) + V_{-,s}(t)$ for a large fixed $t \gg 1/\swaprate$, rescaled by $\sqrt{\transDiffusion t}$ as a function of $\Pe$ as a sanity check for the asymptotic result in Eq.~\eqref{eq:asymptotic_volume_explored}. For $\swaprate t \gg 1$, i.e. when the RnT particle is expected to have changed its orientation many times, it behaves essentially like a diffusive particle and its position is essentially Gaussian \cite{malakarSteadyStateRelaxation2018a}. One may wonder whether the long time limit, $t \gg 1/\swaprate$, gives the self-propulsion enough time to deviate the total volume explored from that of a Brownian particle through long, record-breaking, ballistic runs. As we briefly show in Appendix \ref{app:Excursion_statistics}, the length of the longest uninterrupted ballistic motion, after $N\equiv \alpha t$ re-orientations, scales like $\ln(\alpha t)$ and is therefore sub-leading in $t$ compared to $t^{1/2}$ in Eq.~\eqref{eq:asymptotic_volume_explored}.

Even when Eq.~\eqref{eq:asymptotic_volume_explored} suggests diffusive scaling, the ``local colouring-in" must be dominated by the ballistic motion, whenever the range $\selfpropulsion/\swaprate$ explored by the RnT particle between re-orientations is large compared to the characteristic distance $\transDiffusion/\selfpropulsion$, i.e. $\Pe \gg 1$, explored diffusively before ballistic motion takes over. Said another way, we expect a right-moving RnT particle to explore more regions to the right of its initial position compared to its left and vice versa for a left-moving particle. To quantify this asymmetry induced by the self-propulsion, we compute the total volume explored on the positive half line:
\begin{equation} \label{eq:volume_explored_defn_half_line}
    V^{+}_{s,s'}(t) \equiv \int_0^{\infty} \dint{x} \visitprob_{s,s'}(x,t)
\end{equation}
where the superscript $+$ corresponds to the volume explored on $x\in [0,\infty)$. To relate $ V^{+}_{s,s'}(t)$ to the Fourier transform of $\visitprob_{s,s'}(x,t)$, we extend the integral in Eq.~\eqref{eq:volume_explored_defn_half_line} to $\Rset$ by introducing a Heaviside $\theta-$function
\begin{align} \label{eq:volume_explored_defn_half_line_fourier}
     V^{+}_{s,s'}(t) &= \int_{-\infty}^{\infty} \dint{x} \visitprob_{s,s'}(x,t) \theta(x) = \lim_{\epsilon \to 0^{+}}\int\dintbar{k} \frac{1}{-\imag k + \epsilon}  \int_{-\infty}^{\infty} \dint{x} \exp{-\imag k x} \visitprob_{s,s'}(x,t) \nonumber \\
     & = \lim_{\epsilon \to 0^{+}}\int\dintbar{k} \frac{\visitprob_{s,s'}(k,t)}{-\imag k +\epsilon},
\end{align}
where we have used the Fourier transform of the Heaviside function. Symmetrising the integral over $k$ by changing variables $k \to -k$, we can simplify Eq.~\eqref{eq:volume_explored_defn_half_line_fourier} further:
\begin{equation}\label{eq:volume_explored_defn_half_line_fourier_simplified}
    V^{+}_{s,s'}(t) = \frac{1}{2} \visitprob_{s,s'}(k= 0,t) + \imag \lim_{\epsilon \to 0^{+}}\int \dintbar{k} \frac{k}{k^2 + \epsilon^2} \left( \frac{\visitprob_{s,s'}(k,t) - \visitprob_{s,s'}(-k,t)}{2}\right)
\end{equation}
where we have used the fact that 
\begin{equation}
    \diracdelta{k} = \frac{1}{\pi}\lim_{\epsilon \to 0^{+}} \frac{\epsilon}{k^2 + \epsilon^2}.
\end{equation}
The first term on the RHS of \Eref{volume_explored_defn_half_line_fourier_simplified},
$\visitprob_{s,s'}(k = 0,t)$, is already computed in Eq. \eqref{eq:visit_prob_fourier} for long times. The integral on the right of \Eref{volume_explored_defn_half_line_fourier_simplified}, depends on the odd part of $\visitprob_{s,s'}(k,t)$, Eq.~\eqref{eq:visit_prob_fourier},
\begin{equation} \label{eq:oddpartvisitprob}
     \frac{\bm{\visitprob}(k,\omega) -  \bm{\visitprob}(-k,\omega)}{2} = \frac{1}{-\imag \omega + \varepsilon} \bm{T}^{-1}(\omega) \cdot \begin{pmatrix}
         1 &0\\
         0 & -1
     \end{pmatrix}
     \frac{-\imag \selfpropulsion k}{(-\imag \omega + \transDiffusion k^2 + \swaprate)(-\imag \omega  + \transDiffusion k^2) + \selfpropulsion^2 k^2}.
\end{equation}
Using \Eref{oddpartvisitprob}, we can evaluate \Eref{volume_explored_defn_half_line_fourier_simplified}, yielding
\begin{multline} \label{eq:fourier_transform_volume_explored_hl}
    \bm{V^{+}}(t) = \begin{pmatrix}
        V^{+}_{++}(t) & V^{+}_{+-}(t)\\
        V^{+}_{-+}(t) &  V^{+}_{--}(t)
    \end{pmatrix}\\=\frac{1}{2}\int \dintbar{\omega} e^{-\imag \omega t}
    \begin{pmatrix}
        \frac{\transDiffusion(\sqrt{\beta_+} + \sqrt{\beta_{-}}) + \selfpropulsion }{-\imag \omega\sqrt{-\imag \omega(-\imag \omega +\swaprate)}} & -\frac{\alpha\left(\transDiffusion(\sqrt{\beta_+} + \sqrt{\beta_{-}}) + \selfpropulsion\right) }{\omega \left(\alpha(2\omega + \imag \sqrt{-\imag \omega(-\imag \omega +\alpha)} ) + 2\omega(-\imag \omega + \sqrt{-\imag \omega(-\imag \omega +\swaprate)}) \right)}  \\
        -\frac{\alpha\left(\transDiffusion(\sqrt{\beta_+} + \sqrt{\beta_{-}}) - \selfpropulsion\right) }{\omega \left(\alpha(2\omega + \imag \sqrt{-\imag \omega(-\imag \omega +\alpha)} ) + 2\omega(-\imag \omega + \sqrt{-\imag \omega(-\imag \omega +\swaprate)}) \right)} & \frac{\transDiffusion(\sqrt{\beta_+} + \sqrt{\beta_{-}}) -\selfpropulsion}{-\imag \omega\sqrt{-\imag \omega(-\imag \omega +\swaprate)}}
    \end{pmatrix}
    .
\end{multline}
Focusing again on the $\omega-$poles of \Eref{fourier_transform_volume_explored_hl} located close to the origin, which govern the long-time behaviour of the volume explored $\bm{V^{+}}(t)$, we can find its asymptotic scaling as, Eq.~\eqref{eq:integral_identity},
\begin{equation} \label{eq:asymptotic_volume_explored_half_line}
     \bm{V^{+}}(t)= \begin{pmatrix}
        V^{+}_{++}(t) & V^{+}_{+-}(t)\\
        V^{+}_{-+}(t) &  V^{+}_{--}(t)
    \end{pmatrix}=\frac{\sqrt{\transDiffusion t}}{2 \Gamma(3/2)} 
     \begin{pmatrix}
         \sqrt{1+\Pe} + \sqrt{\Pe} & \sqrt{1 + \Pe} + \sqrt{\Pe} \\
         \sqrt{1+ \Pe} - \sqrt{\Pe} & \sqrt{1+ \Pe} - \sqrt{\Pe}
     \end{pmatrix}
     +\mathcal{O}(t^0).
\end{equation}
$V^{+}_{s,s'}(t)$ is the total volume visited for the first time by time $t$ in state s on the positive half-line by a particle initialised at the origin at time $t=0$ and in state $s'$. Similar to a Brownian particle, this scales like $t^{1/2}$ asymptotically, but due to the self-propulsion, it is not simply the half of the total volume explored $V_{s,s'}(t)$, \Eref{asymptotic_volume_explored}. As expected, irrespective of the initial state, there are more right tracer particles on the positive half line as $\sqrt{1+\Pe} + \sqrt{\Pe} \geq \sqrt{1+\Pe} - \sqrt{\Pe}$ for all Péclet numbers $\Pe\geq 0$. Furthermore, in the limit $\Pe \to \infty$, we do not find any left tracers on the right half line as a particle cannot visit \emph{for the first time} a position on the right half line as a left mover. The ratio of ``diffusive puddle'' over ``ballistic pond'', 
\begin{equation}
    \frac{V^{+}_{-,s'}(t)}{V^{+}_{+,s'}(t)} = \frac{\sqrt{1+\Pe} - \sqrt{\Pe}}{\sqrt{1+\Pe} + \sqrt{\Pe}} + \mathcal{O}(t^{-1/2}) = \left(\sqrt{1+\Pe} - \sqrt{\Pe} \right)^2 + \mathcal{O}(t^{-1/2})
\end{equation}
vanishes correspondingly with divergent Péclet number, $\Pe \to \infty$, for long times and attains its maximum of unity in the Brownian case. As a sanity check, we compare our results in \Erefs{asymptotic_volume_explored} and \eqref{eq:asymptotic_volume_explored_half_line} with simulations illustrated in Fig. \ref{fig:simul_comparison}. 

\section{Conclusion}\label{sec:conc}
In conclusion, we have used Doi-Peliti field theory to characterise the state-dependent visit probability of a RnT particle, namely the probability density that a RnT particle visits a position $x$ for the first time as a right (or left) mover by time $t$ given it was initialised as a right (or left) particle at position $x_0$ at time $t_0$. Unlike the canonical definitions of survival probability \cite{Cox1970}, the state-dependent visit probability incorporates the internal degree of freedom of the RnT particle, allowing for a more comprehensive characterisation of its extreme value statistics. 

We have extended the tracer mechanism of \cite{walterFieldTheorySurvival2022, nekovarFieldtheoreticApproachWiener2016} by endowing the immobile tracer particles with an internal degree of freedom reflecting the self-propulsion state of their parent particle, which allows for a field-theoretic treatment of the problem. We have fully calculated the state-dependent visit probability, \Erefs{visit_prob_fourier} and (\ref{eq:volume_def}) of the particle in Fourier space, from which we extracted the asymptotic behaviour, \Eref{asymptotic_volume_explored}, allowing us to calculate the volume explored by a particle of a certain orientation. 
We have shown that the volume explored by a RnT particle asymptotically scales like $t^{1/2}$, corresponding to the Brownian result. The prefactor of the power-law scaling also shows similarity with the Brownian result with the diffusion constant, $\transDiffusion$, being replaced by the effective diffusion constant of the particle $D_{\rm eff} = \transDiffusion + \selfpropulsion^2/\swaprate$. Finally, we have also characterised the asymmetry of the volume explored by calculating the volume explored by a particle of a certain orientation on the positive half line, \Eref{asymptotic_volume_explored_half_line}. 

As expected, for any non-vanishing self-propulsion, positions to the right of the particle's starting point are visited for the first time predominantly when the particle is in the right-mover state. This effect is strongest when the translational diffusion vanishes, whereby an RnT particle can visit any position on the positive half line for the first time only as a right mover.

Our results demonstrate the robustness of Doi-Peliti field theory in characterising extreme value statistics of particle dynamics. Characterising \emph{state-dependent} statistics both deepen our understanding of paradigmatic active matter models and enable more efficient control of active matter systems. Indeed, these statistics have proven to be essential in creating efficient active information engines \cite{sezik2025conditionalsplittingprobabilitieshiddenstate,szilard_chen,szilard_chen, PhysRevLett.131.188301}. Our results can be straightforwardly extended to calculate state-dependent mean-first passage time. 

One further interesting observable would be the \emph{state-dependent} splitting probabilities, quantifying the mutually exclusive probability that a particle leaving a bounded interval through one exit rather than the other. Such statistics have been drawn on \cite{PhysRevLett.131.188301} in the context of a boundary-update protocol to extract work from RnT particles whose instantaneous orientation is hidden. They have also been argued to be useful in inferring the hidden internal states of active particles \cite{sezik2025conditionalsplittingprobabilitieshiddenstate}. Computation of such statistics normally require the implementation of boundary conditions, breaking the spatial translational symmetry and thereby spoiling a lot of the structure used in this work. The present framework might allow an alternative implementation using correlation functions. Characterising these statistics represent an interesting problem that is left for future work. 
Furthermore, having constructed a field theory, we may allow for interactions between the particle and tracers, reminiscent of a self-interacting random walk \cite{Bremont2024Exact, Bremont2025Persistence}. In endowing the system with such an interaction, one can approximate a biological system of agents modifying their environment with polar cues that promote biased motion.

\section*{Acknowledgments}
The authors would like to thank Julien Br\'{e}mont, Paul Pineau and Jos\'{e} Giral-Barajas for interesting conversations. E.S. and C.B. were supported by Roth PhD scholarships funded by the Department of Mathematics at Imperial College London.
\appendix


\section{Scaling of longest excursion} \label{app:Excursion_statistics}
In this section, we calculate the average distance covered by an RnT particle during its longest ballistic excursion. This is to confirm that an RnT particle does not experience increasingly long ballistic excursions, during which it would cover the real line faster than $\sim \sqrt{t}$, as one might speculate. For this calculation, we fix the number of tumbles $N$ that change the particle's orientation and assume that the particle was initialised at time $t_0 = 0$. 

$P_N(\tau)$ denotes the probability density of $\tau$ being the longest uninterrupted time of the particle being in the same self-propulsion state among $N$ state switches which occur with rate $\alpha/2$. The waiting time distribution between two tumbles, $\psi(t)$, is an exponential distribution with intensity $\alpha/2$. For $\tau$ to be the longest excursion, any one of the $N$ "waiting times" must terminate at $\tau$, which happens with density $\alpha/2 \exp{-\alpha \tau/2}$, and $N-1$ others need to be shorter which happens with probability $\left( 1 - \exp{-\swaprate \tau/2}\right)^{N-1}$,
\begin{equation} \label{eq:longest_excursion}
    P_{N}(\tau) = N\left( 1 - \exp{-\swaprate \tau/2}\right)^{N-1} \frac{\swaprate}{2}\exp{-\swaprate\tau/2}.
\end{equation}
The expected time of the longest excursion can now be calculated to be
\begin{equation} \label{eq:expected_longest_excursion}
    \av{\tau} = \int_{0}^{\infty}\dint{\tau} \tau P_{N}(\tau) = \frac{2}{\alpha} N\int_{0}^{\infty} \dint{u} u \exp{-u}(1-\exp{-u})^{N-1} = \frac{2}{\alpha}H(N)
\end{equation}
where $H(N)$ is the $N^{\rm th}$ Harmonic number
\begin{equation} \label{eq:harmonic_number}
    H(N) \equiv \sum_{n = 1}^{N}\frac{1}{n},
\end{equation}
so that \cite{Gradshteyn:2007}
\begin{equation}
    \av{\tau} \approx \frac{2}{\alpha}\left( \ln(N) + \gamma + \mathcal{O}(1/N)\right)
\end{equation}
where $\gamma$ is the Euler-Mascheroni constant. As the longest ballistic excursion $\av{\tau}$ scales only like $\ln(N)$ and $N$ follows a Poisson distribution, the longest stretch by an RnT particle in a single uninterrupted ballistic run up until time $t$ is $\sim \selfpropulsion 2/\alpha \ln(\alpha t)$ and therefore subleading to the diffusive exploration.

\section{Limiting cases} \label{app:limiting_cases}
In this section, we consider different limits of Eq.~\eqref{eq:volume_explored_fourier} where the Fourier transforms can be computed explicitly and exact results for all times can be obtained, either as to compare to the results in the literature or to extend the results above, in particular \Erefs{asymptotic_volume_explored} and (\ref{eq:asymptotic_volume_explored_half_line}). There are four such cases.

\subsection{$\selfpropulsion\to0, \swaprate\to0$}
For $\selfpropulsion\to0$, in addition to $\swaprate\to0$, the RnT particle becomes a passive Brownian particle. In this limit, the orientation of the particle does not evolve and thus is determined by the initialisation. We expect to recover the standard result of a Brownian particle, essentially the Wiener sausage problem in one dimension \cite{nekovarFieldtheoreticApproachWiener2016,berezhkovskii1989wiener}. 

Taking the limits, Eq.~\eqref{eq:volume_explored_fourier} becomes:
\begin{equation} \label{eq:volume_explored_fourier_limit_1}
    \visitprob_{s,s'}(k = 0,\omega) = \frac{2\sqrt{\transDiffusion} }{(-\imag \omega + \epsilon)^{3/2}} \delta_{s,s'},
\end{equation}
as $\beta_{\pm} = -\imag \omega/ \transDiffusion$, \Eref{betafactordefn}.
Fourier transforming Eq.~\eqref{eq:volume_explored_fourier_limit_1} using the integral identity in Eq.~\eqref{eq:integral_identity}, we find
\begin{equation} \label{eq:volume_explored_limit_1}
    V_{s,s'}(t) = \frac{2\sqrt{\transDiffusion t}}{\Gamma(3/2)}\delta_{s,s'} = 4\sqrt{\frac{\transDiffusion t}{\pi}}\delta_{s,s'}
\end{equation}
While Eq.~\eqref{eq:volume_explored_limit_1} matches the result of the volume explored by a Brownian point-particle \cite{nekovarFieldtheoreticApproachWiener2016, berezhkovskii1989wiener} and it agrees with \Eref{total_Volume_covered} for $\Pe = 0$, it differs from \Eref{asymptotic_volume_explored} by a factor of $2$. This is the result of $\alpha \to 0$ limit being taken before considering $\alpha t \gg 1$ as in \Eref{asymptotic_volume_explored}

\subsection{$\transDiffusion\to0^{+}, \swaprate\to0$}
For $\transDiffusion\to0^{+}, \swaprate\to0$, the RnT moves purely ballistically with the orientation specified by the initialisation. Once again, the orientation of the particle does not evolve. In such a case, we expect to recover a ballistic scaling of the volume explored, $V_{s,s'}(t) = \selfpropulsion t \delta_{s,s'}$.

Taking these limits, Eq.~\eqref{eq:volume_explored_fourier} becomes:
\begin{equation} \label{eq:volume_explored_fourier_limit_2}
    \visitprob_{s,s'}(k = 0,\omega) = \frac{\selfpropulsion }{(-\imag \omega + \epsilon)^{2}} \delta_{s,s'}
\end{equation}
as $\transDiffusion \sqrt{\beta_{+}} \to \selfpropulsion$ and $\transDiffusion \sqrt{\beta_{-}} \to 0$, \Eref{betafactordefn}. Fourier transforming Eq.~\eqref{eq:volume_explored_fourier_limit_2} according to the integral identity in Eq.~\eqref{eq:integral_identity}, we find,
\begin{equation} \label{eq:volume_explored_limit_2}
    V_{s,s'}(t) = \selfpropulsion t \delta_{s,s'},
\end{equation}
as expected. \Eref{volume_explored_limit_2} cannot be recovered by taking an appropriate limit of \Eref{total_Volume_covered} as the latter was obtained under the assumption $\swaprate > 0$. Taking $\transDiffusion \to 0^{+}$, $\swaprate \to 0$ in \Eref{oddpartvisitprob} and using Eq.~\eqref{eq:volume_explored_defn_half_line_fourier_simplified} with \Eref{volume_explored_limit_2}, the volume explored on the positive half-line becomes
\begin{equation} \label{eq:volume_explored_half_line_limit_2}
    V^{+}_{s,s'}(t) = \selfpropulsion t \delta_{s,+} \delta_{s',+},
\end{equation}
which is $\selfpropulsion t$ only when we probe for a right-mover on the positive half-line, after initialising a particle as a right-mover. Otherwise, the range explored vanishes as the particle moves ballistically to the left and never tumbles. 

\subsection{$\selfpropulsion\to0$, $\swaprate >0$}
For $\selfpropulsion\to0$ and $\swaprate\to0$, the RnT particle becomes a passive Brownian particle endowed with an internal degree of freedom that undergoes a telegraph process. The orientation of the particle evolves but has no bearing on the dynamics of the particle. In this case, we expect the total range explored by the particle, $\visitprob_{+,s'} + \visitprob_{-,s'}$, to be given by \Eref{volume_explored_fourier_limit_2}, which covers the case of a Brownian particle whose initial state is frozen.

Taking $\selfpropulsion \to 0$, while keeping $\swaprate$ finite, Eq.~\eqref{eq:volume_explored_fourier} becomes:
\begin{equation} \label{eq:volume_explored_fourier_limit_3}
    \visitprob_{s,s'}(k = 0,\omega) = \sqrt{\transDiffusion}
    \begin{pmatrix}
        \frac{\sqrt{-\imag \omega +\alpha} + \sqrt{-\imag \omega}}{-\imag \omega\sqrt{-\imag \omega(-\imag \omega +\swaprate)}} & -\frac{\alpha(\sqrt{-\imag \omega +\alpha} + \sqrt{-\imag \omega})}{\omega \left(\alpha(2\omega + \imag \sqrt{-\imag \omega(-\imag \omega +\alpha)} ) + 2\omega(-\imag \omega + \sqrt{-\imag \omega(-\imag \omega +\swaprate)}) \right)}  \\
        -\frac{\alpha(\sqrt{-\imag \omega +\alpha} + \sqrt{-\imag \omega}) }{\omega \left(\alpha(2\omega + \imag \sqrt{-\imag \omega(-\imag \omega +\alpha)} ) + 2\omega(-\imag \omega + \sqrt{-\imag \omega(-\imag \omega +\swaprate)}) \right)} & \frac{\sqrt{-\imag \omega +\alpha} + \sqrt{-\imag \omega}}{-\imag \omega\sqrt{-\imag \omega(-\imag \omega +\swaprate)}}
    \end{pmatrix}
\end{equation}
as $\beta_{\pm} = (-\imag \omega + \swaprate/2 \pm \swaprate/2)/\transDiffusion$, \Eref{betafactordefn}. As expected, summing over the columns or, in fact, rows we find:
\begin{equation}
    \sum_{s' \in \{+,-\}} \visitprob_{s,s'}(k = 0,\omega) = \frac{2 \sqrt{\transDiffusion}}{(-\imag \omega + \epsilon)^{3/2}}
\end{equation}
whereby the total volume covered by the particle, \Eref{volume_def} becomes the Brownian result
\begin{equation} \label{eq:total_volume_explored_limit_3}
    \sum_{s \in \{+,-\}} V_{s,s'}(t) = 4\sqrt{\frac{\transDiffusion t}{\pi}}
\end{equation}
independent of the initial orientations. Having calculated the sum $V_{+,s'}(t) + V_{-,s'}(t)$, we now proceed to determine $V_{+,+}(t) = V_{-,-}(t)$, directly from \Eref{volume_explored_fourier_limit_3} and derive $V_{+,-}(t) = V_{-,+}(t)$ through \Eref{total_volume_explored_limit_3}. Regularising again, the diagonal elements of \Eref{volume_explored_fourier_limit_3} give
\begin{equation}
    V_{++}(t) = V_{--}(t) = \sqrt{\transDiffusion} \int \dintbar{\omega} \exp{-\imag \omega t}\left[ \frac{1}{(-\imag \omega + \epsilon)^{3/2}} + \frac{1}{(-\imag \omega + \epsilon)\sqrt{-\imag \omega +\alpha} }\right].
\end{equation}
Using Eq.~\eqref{eq:integral_identity} and the Schwinger trick, $(-\imag \omega + \epsilon)^{-1/2} = \int_0^{\infty}\dint{s} \exp{-s(-\imag \omega +\epsilon)}/\sqrt{\pi s}$, these Fourier transforms are evaluated to give:
\begin{equation} \label{eq:volume_explored_diag_limit_3}
    V_{++}(t) = V_{--}(t) = 2\sqrt{\frac{\transDiffusion t}{\pi}}\left( 1 + \frac{\gamma(1/2, \alpha t)}{2\sqrt{\alpha t}}\right),
\end{equation}
where $\gamma(n,x) \corresponding \int_{0}^{x} \dint{s} s^{n-1} \exp{-s}$, is the incomplete gamma function. Subtracting \Eref{volume_explored_diag_limit_3} from \Eref{total_volume_explored_limit_3}, the off-diagonal elements can be found
\begin{equation}
     V_{+-}(t) = V_{-+}(t) = 2\sqrt{\frac{\transDiffusion t}{\pi}}\left( 1 - \frac{\gamma(1/2, \alpha t)}{2\sqrt{\alpha t}}\right).
\end{equation}
Finally, since $\visitprob_{s,s'}(k,\omega)$ is even in $k$ for $\selfpropulsion  = 0$, \Erefs{volume_def} and \Eref{volume_explored_defn_half_line_fourier_simplified} give
\begin{equation}
    V^{+}_{s,s'}(t) = \frac{1}{2}V_{s,s'}(t)
\end{equation}

\subsection{$\transDiffusion\to0^{+}$, $\swaprate > 0$}
Finally, we consider the limit $\transDiffusion \to 0^{+}$ with $\swaprate$ being kept finite. The RnT particle no longer diffuses but moves ballistically in a certain direction before changing its orientation with rate $\alpha/2$. Without the diffusion, all points to the right of the particle's initial position are visited for the first time only as a right mover and those to the left only as a left-mover.

Taking $\transDiffusion \to 0^{+}$, Eq.~\eqref{eq:volume_explored_fourier} becomes with $\lim_{\transDiffusion \to 0^{+}} \transDiffusion(\sqrt{\beta_{+}} + \sqrt{\beta_{-}}) = \selfpropulsion$
\begin{equation} \label{eq:volume_explored_fourier_limit_4}
    \visitprob_{s,s'}(k = 0,\omega) = \selfpropulsion
    \begin{pmatrix}
        \frac{1}{-\imag \omega\sqrt{-\imag \omega(-\imag \omega +\swaprate)}} & -\frac{\alpha}{\omega \left(\alpha(2\omega + \imag \sqrt{-\imag \omega(-\imag \omega +\alpha)} ) + 2\omega(-\imag \omega + \sqrt{-\imag \omega(-\imag \omega +\swaprate)}) \right)}  \\
        -\frac{\alpha}{\omega \left(\alpha(2\omega + \imag \sqrt{-\imag \omega(-\imag \omega +\alpha)} ) + 2\omega(-\imag \omega + \sqrt{-\imag \omega(-\imag \omega +\swaprate)}) \right)} & \frac{1}{-\imag \omega\sqrt{-\imag \omega(-\imag \omega +\swaprate)}}
    \end{pmatrix}
    .
\end{equation}
Only the diagonal elements can be Fourier transformed in closed form. Using Schwinger's trick and Eq.~\eqref{eq:integral_identity}, we find:
\begin{equation}
    V_{++}(t) = V_{--}(t) = \selfpropulsion t \exp{-\swaprate t/2}\left[I_{0}\left( \frac{\alpha t}{2} \right) + I_{1}\left( \frac{\alpha t}{2} \right) \right]
\end{equation}
where $I_{n}(x)$ is the Modified Bessel function of the first kind \cite{mathematica}
\begin{equation}
    \int_{0}^{1}\dint{y} \sqrt{\frac{1-y}{y}} \exp{-y \swaprate t} = \frac{\pi}{2}\exp{-\swaprate t/2}\left[I_{0}\left( \frac{\alpha t}{2} \right) + I_{1}\left( \frac{\alpha t}{2} \right) \right].
\end{equation}
For $\alpha \to 0$, we recover Eq.~\eqref{eq:volume_explored_limit_2}, as $I_{0}(0) = 1$ and $I_{1}(0) = 0$, and for $t \gg \alpha$, we recover the result in Eq.~\eqref{eq:asymptotic_volume_explored}, as $I_{0,1}(z) \sim \exp{z}/\sqrt{2\pi z}$. Finally, in the limit of $\transDiffusion \to 0$ for finite $\alpha>0$, the volume explored on the positive half-line for $\selfpropulsion > 0$ is given by
\begin{equation}
    V^{+}_{s,s'}(t) = \selfpropulsion\int \dintbar{\omega} \exp{-\imag \omega t}
    \begin{pmatrix}
        \frac{1}{-\imag \omega\sqrt{-\imag \omega(-\imag \omega +\swaprate)}} & -\frac{\alpha}{\omega \left(\alpha(2\omega + \imag \sqrt{-\imag \omega(-\imag \omega +\alpha)} ) + 2\omega(-\imag \omega + \sqrt{-\imag \omega(-\imag \omega +\swaprate)}) \right)}  \\
       0 & 0
    \end{pmatrix}
\end{equation}
where the bottom row is $0$ because a left-moving particle can never visit for the first time a point on the positive half line and $V^{+}_{+,s'}(t) = V_{+,s}(t)$ since right tracers can only be found in the positive half-line. This is equally obtained by taking the limit $\transDiffusion \to 0^{+}$ in \Eref{volume_explored_defn_half_line_fourier_simplified} using \Erefs{oddpartvisitprob} and (\ref{eq:volume_explored_fourier_limit_4}).
\bibliography{RnT_skidmarks}

\end{document}